\documentclass[fleqn, usenatbib]{mnras}

\usepackage{newtxtext,newtxmath}

\usepackage[T1]{fontenc}
\usepackage[utf8]{inputenc}
\usepackage[british]{babel}

\usepackage{amsmath}
\usepackage{mathtools}
\usepackage{graphicx}
\usepackage{xcolor}
\usepackage{xparse}
\usepackage{mleftright}
\usepackage{xfrac}
\usepackage{microtype}
\usepackage{newunicodechar}

\usepackage{silence}
\ActivateWarningFilters[pdflatex-unicode-math]

\newunicodechar{†}{\dag}
\newunicodechar{‡}{\ddag}
\newunicodechar{…}{\ldots}
\newunicodechar{⋯}{\cdots}
\newunicodechar{⋮}{\vdots}
\newunicodechar{⋱}{\ddots}
\newunicodechar{⋰}{\iddots}
\newunicodechar{α}{\alpha}
\newunicodechar{β}{\beta}
\newunicodechar{γ}{\gamma}
\newunicodechar{δ}{\delta}
\newunicodechar{ε}{\varepsilon}
\newunicodechar{ϵ}{\epsilon}
\newunicodechar{ζ}{\zeta}
\newunicodechar{η}{\eta}
\newunicodechar{θ}{\theta}
\newunicodechar{ϑ}{\vartheta}
\newunicodechar{ι}{\iota}
\newunicodechar{κ}{\kappa}
\newunicodechar{ϰ}{\varkappa}
\newunicodechar{λ}{\lambda}
\newunicodechar{μ}{\mu}
\newunicodechar{ν}{\nu}
\newunicodechar{ξ}{\xi}
\newunicodechar{ο}{o}
\newunicodechar{π}{\pi}
\newunicodechar{ρ}{\rho}
\newunicodechar{ϱ}{\varrho}
\newunicodechar{σ}{\sigma}
\newunicodechar{τ}{\tau}
\newunicodechar{υ}{\upsilon}
\newunicodechar{φ}{\varphi}
\newunicodechar{ϕ}{\phi}
\newunicodechar{χ}{\chi}
\newunicodechar{ψ}{\psi}
\newunicodechar{ω}{\omega}
\newunicodechar{Α}{\mathrm{A}}
\newunicodechar{Β}{\mathrm{B}}
\newunicodechar{Γ}{\Gamma}
\newunicodechar{Δ}{\Delta}
\newunicodechar{Ε}{\mathrm{E}}
\newunicodechar{Ζ}{\mathrm{Z}}
\newunicodechar{Η}{\mathrm{H}}
\newunicodechar{Θ}{\Theta}
\newunicodechar{Ι}{\mathrm{I}}
\newunicodechar{Κ}{\mathrm{K}}
\newunicodechar{Λ}{\Lambda}
\newunicodechar{Μ}{\mathrm{M}}
\newunicodechar{Ν}{\mathrm{N}}
\newunicodechar{Ξ}{\Xi}
\newunicodechar{Ο}{\mathrm{O}}
\newunicodechar{Π}{\Pi}
\newunicodechar{Ρ}{\mathrm{P}}
\newunicodechar{Σ}{\Sigma}
\newunicodechar{Τ}{\mathrm{T}}
\newunicodechar{Υ}{\Upsilon}
\newunicodechar{Φ}{\Phi}
\newunicodechar{Χ}{\Chi}
\newunicodechar{Ψ}{\Psi}
\newunicodechar{Ω}{\Omega}
\newunicodechar{∑}{\sum}
\newunicodechar{∫}{\int}
\newunicodechar{∬}{\iint}
\newunicodechar{∭}{\iiint}
\newunicodechar{⨌}{\iiiint}
\newunicodechar{∮}{\oint}
\newunicodechar{∯}{\oiint}
\newunicodechar{∰}{\oiiint}
\newunicodechar{∇}{\nabla}
\newunicodechar{∂}{\partial}
\newunicodechar{√}{\sqrt}
\newunicodechar{∈}{\in}
\newunicodechar{∋}{\ni}
\newunicodechar{∉}{\notin}
\newunicodechar{∀}{\forall}
\newunicodechar{∃}{\exists}
\newunicodechar{∄}{\nexists}
\newunicodechar{∴}{\therefore}
\newunicodechar{∵}{\because}
\newunicodechar{⟨}{\langle}
\newunicodechar{⟩}{\rangle}
\newunicodechar{〈}{\langle}
\newunicodechar{〉}{\rangle}
\newunicodechar{⌊}{\lfloor}
\newunicodechar{⌋}{\rfloor}
\newunicodechar{⌈}{\lceil}
\newunicodechar{⌉}{\rceil}
\newunicodechar{∼}{\sim}
\newunicodechar{∝}{\propto}
\newunicodechar{∞}{\infty}
\newunicodechar{ℵ}{\aleph}
\newunicodechar{ℏ}{\hbar}
\newunicodechar{℘}{\wp}
\newunicodechar{ℓ}{\ell}
\newunicodechar{∅}{\emptyset}
\newunicodechar{×}{\times}
\newunicodechar{⋅}{\cdot}
\newunicodechar{÷}{\div}
\newunicodechar{⋆}{\star}
\newunicodechar{∘}{\circ}
\newunicodechar{⋄}{\diamond}
\newunicodechar{⊕}{\oplus}
\newunicodechar{⊖}{\ominus}
\newunicodechar{⊗}{\otimes}
\newunicodechar{⊘}{\oslash}
\newunicodechar{⊙}{\odot}
\newunicodechar{±}{\pm}
\newunicodechar{∓}{\mp}
\newunicodechar{≈}{\approx}
\newunicodechar{≃}{\simeq}
\newunicodechar{≡}{\equiv}
\newunicodechar{≠}{\ne}
\newunicodechar{≥}{\ge}
\newunicodechar{≤}{\le}
\newunicodechar{≳}{\gtrsim}
\newunicodechar{≲}{\lesssim}
\newunicodechar{≫}{\gg}
\newunicodechar{≪}{\ll}
\newunicodechar{⊂}{\subset}
\newunicodechar{⊃}{\supset}
\newunicodechar{⊆}{\subseteq}
\newunicodechar{⊇}{\supseteq}
\newunicodechar{⊈}{\nsubseteq}
\newunicodechar{⊉}{\nsupseteq}
\newunicodechar{≔}{\coloneqq}
\newunicodechar{≕}{\eqqcolon}
\newunicodechar{¬}{\neg}
\newunicodechar{∨}{\vee}
\newunicodechar{∧}{\wedge}
\newunicodechar{∪}{\cup}
\newunicodechar{∩}{\cap}
\newunicodechar{⋁}{\bigvee}
\newunicodechar{⋀}{\bigwedge}
\newunicodechar{⋃}{\bigcup}
\newunicodechar{⋂}{\bigcap}
\newunicodechar{⟂}{\perp}
\newunicodechar{∥}{\parallel}
\newunicodechar{∦}{\nparallel}
\newunicodechar{𝚤}{\imath}
\newunicodechar{𝚥}{\jmath}
\newunicodechar{⇔}{\Leftrightarrow}
\newunicodechar{⇕}{\Updownarrow}
\newunicodechar{⇐}{\Leftarrow}
\newunicodechar{⇒}{\Rightarrow}
\newunicodechar{⇑}{\Uparrow}
\newunicodechar{⇓}{\Downarrow}
\newunicodechar{↔}{\leftrightarrow}
\newunicodechar{↕}{\updownarrow}
\newunicodechar{←}{\leftarrow}
\newunicodechar{→}{\rightarrow}
\newunicodechar{↑}{\uparrow}
\newunicodechar{↓}{\downarrow}
\newunicodechar{⟷}{\longleftrightarrow}
\newunicodechar{⟵}{\longleftarrow}
\newunicodechar{⟶}{\longrightarrow}
\newunicodechar{⇇}{\leftleftarrows}
\newunicodechar{⇉}{\rightrightarrows}
\newunicodechar{⇈}{\upuparrows}
\newunicodechar{⇊}{\downdownarrows}
\newunicodechar{⟺}{\Longleftrightarrow}
\newunicodechar{⟸}{\Longleftarrow}
\newunicodechar{⟹}{\Longrightarrow}
\newunicodechar{↦}{\mapsto}
\newunicodechar{↤}{\mapsfrom}
\newunicodechar{⟼}{\longmapsto}
\newunicodechar{⟻}{\longmapsfrom}
\newunicodechar{⟾}{\Longmapsto}
\newunicodechar{⟽}{\Longmapsfrom}
\newunicodechar{↗}{\nearrow}
\newunicodechar{↖}{\nwarrow}
\newunicodechar{↘}{\searrow}
\newunicodechar{↙}{\swarrow}
\newunicodechar{↩}{\hookleftarrow}
\newunicodechar{↪}{\hookrightarrow}
\newunicodechar{↶}{\curvearrowleft}
\newunicodechar{↷}{\curvearrowright}
\newunicodechar{↺}{\circlearrowleft}
\newunicodechar{↻}{\circlearrowright}
\newunicodechar{↫}{\looparrowleft}
\newunicodechar{↬}{\looparrowright}
\newunicodechar{⇋}{\leftrightharpoons}
\newunicodechar{⇌}{\rightleftharpoons}
\newunicodechar{↼}{\leftharpoonup}
\newunicodechar{↽}{\leftharpoondown}
\newunicodechar{⇀}{\rightharpoonup}
\newunicodechar{⇁}{\rightharpoondown}
\newunicodechar{↿}{\upharpoonleft}
\newunicodechar{↾}{\upharpoonright}
\newunicodechar{⇃}{\downharpoonleft}
\newunicodechar{⇂}{\downharpoonright}
\newunicodechar{𝔸}{\mathbb{A}}
\newunicodechar{𝔹}{\mathbb{B}}
\newunicodechar{ℂ}{\mathbb{C}}
\newunicodechar{𝔻}{\mathbb{D}}
\newunicodechar{𝔼}{\mathbb{E}}
\newunicodechar{𝔽}{\mathbb{F}}
\newunicodechar{𝔾}{\mathbb{G}}
\newunicodechar{ℍ}{\mathbb{H}}
\newunicodechar{𝕀}{\mathbb{I}}
\newunicodechar{𝕁}{\mathbb{J}}
\newunicodechar{𝕂}{\mathbb{K}}
\newunicodechar{𝕃}{\mathbb{L}}
\newunicodechar{𝕄}{\mathbb{M}}
\newunicodechar{ℕ}{\mathbb{N}}
\newunicodechar{𝕆}{\mathbb{O}}
\newunicodechar{ℙ}{\mathbb{P}}
\newunicodechar{ℚ}{\mathbb{Q}}
\newunicodechar{ℝ}{\mathbb{R}}
\newunicodechar{𝕊}{\mathbb{S}}
\newunicodechar{𝕋}{\mathbb{T}}
\newunicodechar{𝕌}{\mathbb{U}}
\newunicodechar{𝕍}{\mathbb{V}}
\newunicodechar{𝕎}{\mathbb{W}}
\newunicodechar{𝕏}{\mathbb{X}}
\newunicodechar{𝕐}{\mathbb{Y}}
\newunicodechar{ℤ}{\mathbb{Z}}
\newunicodechar{𝒜}{\mathcal{A}}
\newunicodechar{ℬ}{\mathcal{B}}
\newunicodechar{𝒞}{\mathcal{C}}
\newunicodechar{𝒟}{\mathcal{D}}
\newunicodechar{ℰ}{\mathcal{E}}
\newunicodechar{ℱ}{\mathcal{F}}
\newunicodechar{𝒢}{\mathcal{G}}
\newunicodechar{ℋ}{\mathcal{H}}
\newunicodechar{ℐ}{\mathcal{I}}
\newunicodechar{𝒥}{\mathcal{J}}
\newunicodechar{𝒦}{\mathcal{K}}
\newunicodechar{ℒ}{\mathcal{L}}
\newunicodechar{ℳ}{\mathcal{M}}
\newunicodechar{𝒩}{\mathcal{N}}
\newunicodechar{𝒪}{\mathcal{O}}
\newunicodechar{𝒫}{\mathcal{P}}
\newunicodechar{𝒬}{\mathcal{Q}}
\newunicodechar{ℛ}{\mathcal{R}}
\newunicodechar{𝒮}{\mathcal{S}}
\newunicodechar{𝒯}{\mathcal{T}}
\newunicodechar{𝒰}{\mathcal{U}}
\newunicodechar{𝒱}{\mathcal{V}}
\newunicodechar{𝒲}{\mathcal{W}}
\newunicodechar{𝒳}{\mathcal{X}}
\newunicodechar{𝒴}{\mathcal{Y}}
\newunicodechar{𝒵}{\mathcal{Z}}
\newunicodechar{𝕬}{\mathfrak{A}}
\newunicodechar{𝕭}{\mathfrak{B}}
\newunicodechar{𝕮}{\mathfrak{C}}
\newunicodechar{𝕯}{\mathfrak{D}}
\newunicodechar{𝕰}{\mathfrak{E}}
\newunicodechar{𝕱}{\mathfrak{F}}
\newunicodechar{𝕲}{\mathfrak{G}}
\newunicodechar{𝕳}{\mathfrak{H}}
\newunicodechar{𝕴}{\mathfrak{I}}
\newunicodechar{𝕵}{\mathfrak{J}}
\newunicodechar{𝕶}{\mathfrak{K}}
\newunicodechar{𝕷}{\mathfrak{L}}
\newunicodechar{𝕸}{\mathfrak{M}}
\newunicodechar{𝕹}{\mathfrak{N}}
\newunicodechar{𝕺}{\mathfrak{O}}
\newunicodechar{𝕻}{\mathfrak{P}}
\newunicodechar{𝕼}{\mathfrak{Q}}
\newunicodechar{𝕽}{\mathfrak{R}}
\newunicodechar{𝕾}{\mathfrak{S}}
\newunicodechar{𝕿}{\mathfrak{T}}
\newunicodechar{𝖀}{\mathfrak{U}}
\newunicodechar{𝖁}{\mathfrak{V}}
\newunicodechar{𝖂}{\mathfrak{W}}
\newunicodechar{𝖃}{\mathfrak{X}}
\newunicodechar{𝖄}{\mathfrak{Y}}
\newunicodechar{𝖅}{\mathfrak{Z}}

\DeactivateWarningFilters[pdflatex-unicode-math]

\usepackage{siunitx}
\usepackage{fnbreak}
\usepackage{booktabs}
\usepackage{gincltex}
\usepackage{grffile}
\usepackage{import}
\usepackage{tikz}
\usepackage{acro}
\usepackage[useregional, en-GB]{datetime2}
\usepackage{csquotes}
\usepackage{cleveref}

\crefname{figure}{Fig.}{Figs.}
\crefname{table}{Table}{Tables}
\renewcommand*{\vec}{\mathbfit}
\newcommand*{\imag}{\mathrm{i}}
\newcommand*{\euler}{\mathrm{e}}
\newcommand*{\Ms}{\ensuremath{\mathrm{M}_\odot}}
\DeclareSIUnit{\Msunit}{\ensuremath{\mathrm{M}_\odot}}
\newcommand*{\textcode}[1]{\textsc{\small #1}}

\newcommand*{\q}{\enquote}
\newcommand*{\email}[1]{\href{mailto:#1}{\texttt{#1}}}

\crefname{algorithm}{step}{steps}%
\creflabelformat{algorithm}{(#2#1#3)}%

\DeclareSIUnit{\pc}{pc}
\DeclareSIUnit{\kpc}{\kilo\pc}
\DeclareSIUnit{\Mpc}{\mega\pc}
\DeclareSIUnit{\c}{\text{\ensuremath{c}}}
\DeclareSIUnit{\hHubble}{\text{\ensuremath{h}}}
\DeclareSIUnit{\CPU}{CPU}

\newcommand*{\ml}{\mleft}
\newcommand*{\mr}{\mright}
\newcommand*{\mtext}[1]{{\operatorfont #1}}

\newcommand*{\com}{{\mtext{c}}}

\let\Re=\relax
\DeclareMathOperator{\Re}{Re}

\DeclareMathOperator{\FFT}{FFT}

\def\diffd{\mathrm{d}}
\DeclareDocumentCommand\differential{ o g d() }{ 
	\IfNoValueTF{#2}{
		\IfNoValueTF{#3}
		{\diffd\IfNoValueTF{#1}{}{^{#1}}}
		{\mathinner{\diffd\IfNoValueTF{#1}{}{^{#1}}\argopen(#3\argclose)}}
	}
	{\mathinner{\diffd\IfNoValueTF{#1}{}{^{#1}}#2} \IfNoValueTF{#3}{}{(#3)}}
}
\DeclareDocumentCommand\dd{}{\differential} 

\newcommand*{\AREPO}{\textcode{AREPO}}
\newcommand*{\NGenIC}{\textcode{N-GenIC}}
\DeclareAcronym{AMR}{short=AMR, long=adaptive mesh refinement}
\DeclareAcronym{ALP}{short=ALP, long=axion-like particle}
\DeclareAcronym{CDM}{short=CDM, long=cold dark matter}
\DeclareAcronym{CiC}{short=CiC, long=cloud-in-cell}
\DeclareAcronym{FDM}{short=FDM, long=fuzzy dark matter}
\DeclareAcronym{FFT}{short=FFT, long=Fast Fourier Transform}
\DeclareAcronym{FFTW}{short=FFTW, long=Fastest Fourier Transform in the West}
\DeclareAcronym{FoF}{short=FoF, long=friends-of-friends}
\DeclareAcronym{FRW}{short=FRW, long=Friedmann–Robertson–Walker}
\DeclareAcronym{HMF}{short=HMF, long=halo mass function}
\DeclareAcronym{IC}{
	short=IC, long=initial condition,
	foreign-plural={}
}
\DeclareAcronym{LCDM}{short=$Λ$CDM, long=cold dark matter and a cosmological constant}
\DeclareAcronym{MPCDF}{short=MPCDF, long=Max Planck Computing and Data Facility}
\DeclareAcronym{NFW}{short=NFW, long=Navarro–Frenk–White}
\DeclareAcronym{QCD}{short=QCD, long=quantum chromodynamics}
\DeclareAcronym{QFT}{short=QFT, long=quantum field theory}
\DeclareAcronym{SPH}{short=SPH, long=smoothed-particle hydrodynamics}
\DeclareAcronym{WIMP}{
	short=WIMP, long=weakly interacting massive particle,
	foreign-plural={}
}

\title[Structure formation in large-volume cosmological FDM simulations: Impact of non-linear dynamics]{Structure formation in large-volume cosmological simulations of
  fuzzy dark matter: Impact of the non-linear dynamics}

\author[Simon May, Volker Springel]{%
	Simon May$^{1}$\thanks{E-mail: \email{simon.may@pitp.ca}},
	Volker Springel$^{1}$
	\\
	$^{1}$Max-Planck-Institut für Astrophysik, Karl-Schwarzschild-Straße 1, 85741 Garching, Germany
}

\date{Accepted XXX. Received YYY; in original form ZZZ}

\pubyear{2021}

\begin{document}

\label{firstpage}
\pagerange{\pageref{firstpage}--\pageref{lastpage}}
\maketitle

\begin{abstract}
  An ultra-light bosonic particle of mass around
  $10^{-22}\,\mathrm{eV}/c^2$ is of special interest as a dark matter candidate, as it both has
  particle physics motivations, and may give rise to notable
  differences in the structures on highly non-linear scales due to
  the manifestation of quantum-physical wave effects on macroscopic
  scales, which could address a number of contentious
  small-scale tensions in the standard cosmological model, $\Lambda$CDM.
  Using a spectral technique, we here discuss simulations of
  such fuzzy dark matter (FDM), including the full non-linear wave dynamics, with a comparatively large dynamic
  range and for larger box sizes than considered previously.
  While the impact of suppressed small-scale power in the initial conditions associated with FDM has been studied before, the characteristic FDM dynamics are often neglected; in our simulations, we instead show the impact of the full non-linear dynamics on physical observables.
  We focus on the evolution of the matter power spectrum, give first results
  for the FDM halo mass function directly based on full FDM simulations, and discuss the computational challenges associated with the FDM equations.
  FDM shows a pronounced suppression of power on small scales relative to cold dark matter (CDM),
  which can be understood as a damping effect due to \q{quantum pressure}.
  In certain regimes, however, the FDM power can exceed that of CDM, which may be interpreted
  as a reflection of order-unity density fluctuations occurring in
  FDM. In the halo mass function, FDM shows a significant abundance reduction below a characteristic mass scale only.
  This could in principle alleviate the need to invoke very strong feedback processes in small galaxies
  to reconcile $\Lambda$CDM with the observed galaxy luminosity
  function, but detailed studies that also include baryons will be
  needed to ultimately judge the viability of FDM.
\end{abstract}



\section{Introduction}

The \q{standard cosmological model} with \ac{LCDM} has been extremely
successful in describing a wide variety of cosmological observations
across a broad range of physical scales
\citep[e.\,g.][]{Frenk2012, Bull:2015stt}.
On small cosmological scales, however, challenges such as the
\q{cusp–core problem}, the \q{missing satellite problem}, or the
\q{too-big-to-fail problem} have sometimes raised questions about the
validity of \ac{LCDM} \citep{2015PNAS..11212249W,DelPopolo:2016emo,BoylanKolchin2011}.
Since physical effects experienced by baryons
can become relevant at these scales, it has proven difficult to
disentangle the apparent discrepancies from baryonic physics. In
addition, even though the cosmological properties of cold dark matter
are constrained extremely well on large scales, its micro-physical
nature is still completely unknown thus far.

In light of these small-scale questions and the enduring lack of any
direct detection of the most well-studied dark matter candidates, in
particular \acp{WIMP}, models based on ultra-light (\q{axion-like})
scalar particles have been gaining interest as alternative dark matter
models. Due to their small masses, quantum effects are expected to
cause interesting wave-like behaviour at small (i.\,e.\ galactic) scales compared to
heavy particles like \acp{WIMP} or compact objects. These effects
have also been proposed to elucidate some of the \q{small-scale problems}
of \ac{CDM}; for example, early numerical simulations
showed that ultra-light scalars form cores in the centres of dark
matter haloes \citep{Schive:2014dra}. Furthermore, light
(pseudo-)scalar particles are a common feature of theories in particle
physics, from the original axion in \ac{QCD} to a plethora of
axion-like particles predicted by unified and early-universe theories
such as string theories \citep{Marsh:2015xka}.

The amount of existing work studying structure formation with \ac{CDM}
utterly dwarfs that of such \ac{FDM}, particularly concerning numerical
simulations that reliably probe the non-linear regime. Computations
have been performed using a variety of approaches and numerical
methods, although many attempts were limited in scope
\citep[][table~1]{Zhang:2018ghp,2021MNRAS.504.2391L}. Correspondingly, our
understanding of structure formation in \ac{FDM} cosmologies is still
comparatively spotty compared to \ac{CDM}, where decades of experience
have led to extremely detailed insights. Apart from the lower level
of research attention, an important reason impeding insight into
\ac{FDM} has been that numerically solving the corresponding equations
of motion incurs very large computational costs – much higher than
those associated with corresponding calculations of \ac{LCDM}. In
particular, while it has been established that, in the limit of large
scales (or large particle mass), the \ac{FDM} equations are equivalent
to the Vlasov–Poisson equations of \ac{CDM}
\citep{2018PhRvD..97h3519M,1993ApJ...416L..71W}, the effects of
\ac{FDM} in (mildly) non-linear regimes of structure formation are
still poorly understood when compared to \ac{CDM}.

Due to the computational requirements, the cosmological volumes
studied in simulations with full \ac{FDM} dynamics have been especially limited \citep{2009ApJ...697..850W,Schive:2014dra,2018PhRvD..98d3509V,2020MNRAS.494.2027M}, which is an issue that this work seeks to improve upon.
In particular, we would like to
carry out simulations that smoothly connect the non-linear state
reached in isolated \ac{FDM} haloes to the still linear large-scale
structure, thereby bridging, in particular, the regime of mildly
non-linear evolution where differences in the temporal evolution
compared to \ac{CDM} can be expected. To this end, we carry out very
large uni-grid \ac{FDM} simulations with a spectral method, which
fully retains the quantum-mechanical effects. Because there is still
a dearth of precision studies of how \ac{FDM} compares to \ac{CDM} for
traditional measures of large-scale structure, we focus our analysis
on central measures of matter clustering, namely the power spectrum
and the halo mass function, and compare to ordinary \ac{LCDM} where
appropriate.

While methods which forego a treatment of the full wave dynamics have been able to conduct simulations with volumes much closer to those attainable using traditional $N$-body and \ac{SPH} approaches for \ac{CDM} \citep{Schive2016,2016PhRvD..94l3523V,2018ApJ...863...73Z,2018MNRAS.478.3935N,2019MNRAS.482.3227N}, these do not capture inherent wave phenomena such as interference effects, which can have a significant impact on the overall evolution at least on small scales \citep{2019PhRvD..99f3509L}, leaving the validity of results obtained in this way unclear in the absence of similar computations solving the fundamental wave equations.
In particular, while all simulations can easily incorporate the impact of the suppressed small-scale power spectrum present with \ac{FDM} in the initial conditions, such methods either lack the wave nature of \ac{FDM} entirely or only approximate it.
Using our simulations presented in this work, we aim to clarify the reliability of such approximative results by explicitly omitting the suppression in the \ac{FDM} initial conditions, starting instead from \q{standard} \ac{LCDM} initial conditions.
In this way, we disentangle the two essential physical differences distinguishing \ac{FDM} from \ac{CDM} in cosmological numerical simulations: the initial conditions and the non-linear dynamics.

This study is structured as follows. In \cref{sec:theory}, we
briefly review the theoretical basis for the equations of motion of
\ac{FDM} models. We then describe our numerical approaches for
cosmological simulations of \ac{FDM} in \cref{sec:numerics}. In
\cref{sec:power-spectrum}, we turn to an analysis of our results
for the matter power spectrum, while in \cref{sec:mass-function}
we report our findings for the halo mass
function. \Cref{sec:profiles} discusses halo profiles, and
\cref{sec:fdm-ics} the differences expected in \ac{FDM} due to
modifications of the initial linear theory spectrum relative to
\ac{LCDM} if this is self-consistently computed. Finally, we give our
conclusions in \cref{sec:conclusions}.

\section{Theoretical background}
\label{sec:theory}

Fundamentally, \acl{FDM} is identical to the well-studied
scalar field dark matter model, which is perhaps also the simplest particle-based
dark matter model at the theoretical level. Such a model is described
by the simple scalar field action
\begin{equation}
	\label{eq:action}
	S = \frac{1}{ℏc^2} ∫\! \dd[4]{x} √{-g} \ml(\frac{1}{2} g^{μν} (∂_μ ϕ) (∂_ν ϕ) - \frac{1}{2} \frac{m^2 c^2}{ℏ^2} ϕ^2 - \frac{λ}{ℏ^2 c^2} ϕ^4\mr)
\end{equation}
with the metric $g^{μν}$ and its determinant $g$, and the real scalar field $ϕ$, its mass $m$ and a coupling strength $λ$.\footnote{%
	$c$ and $ℏ$ have the usual meanings of the speed of light in vacuum and the reduced Planck constant, and are explicitly included in all equations.%
}
This action is to be understood in the context of quantum field theory in a curved spacetime, i.\,e.\ $ϕ$ is a (\q{second-quantised}) quantum field.

The difference to most earlier studies of scalar field dark matter lies in the chosen parameters, specifically the mass $m$, which is commonly assumed to be in the range of \SI{100}{\GeV}–\si{\TeV}, in line with the concept of \ac{CDM}.
In contrast, an \emph{ultra-light} scalar field with a mass around $m c^2 ≈ \SI{e-22}{\eV}$ is under consideration here.
This vast difference in the considered mass range compared to heavy scalar particles has drastic phenomenological consequences.
Additionally, the ultra-light particles require a \emph{non-thermal} production mechanism so that the resulting dark matter is not ultra-hot, but still resembles \ac{CDM}.

The term \q{\acl{FDM}}, then, is typically used for an
ultra-light scalar field \emph{without} self-interactions, i.\,e.\
$λ = 0$. Thus, it corresponds to a limit ($λ → 0$, or zero
temperature, $T → 0$) of more general ultra-light scalar field models
\citep[see][for a review]{Ferreira2020}, such as superfluid dark
matter. Such models are often called \ac{ALP} models due to the
phenomenological similarity to the axion of \ac{QCD}, which yields the
same action as \cref{eq:action} originating from a periodic potential
$V(ϕ) ∝ Λ^4 (1 - \cos(\sfrac{ϕ}{f_a}))$ for $ϕ ≪ f_a$ \citep[see e.\,g.][]{Marsh:2015xka,Hui:2016ltb}.

In order to perform numerical simulations of the non-linear structure formation in a universe with \ac{FDM}, as they are traditionally done for \ac{CDM}, it is necessary to use non-relativistic approximations.
This is possible because virial velocities are small compared to $c$ and the simulations will be restricted to structures forming on scales smaller than the Hubble horizon $c / H_0$.
Even so, as will be detailed later, \ac{FDM} simulations are only possible with massively increased computational cost compared to \ac{CDM}.

As sketched, for example, in \citet{Marsh:2015xka,Hui:2016ltb}, the non-relativistic equations of motions for the fuzzy dark matter field can be obtained as follows.
First, the real scalar field is rewritten in terms of a complex field $ψ$ as
\begin{equation}
	ϕ
	= \frac{1}{2} √{\frac{ℏ^3 c}{2m}} \Re\ml(ψ \euler^{-\imag\frac{mc^2}{ℏ} t}\mr)
	= √{\frac{ℏ^3 c}{2m}} \ml(ψ \euler^{-\imag\frac{mc^2}{ℏ} t} + ψ^* \euler^{\imag\frac{mc^2}{ℏ} t}\mr).
\end{equation}
Next, the non-relativistic limit is taken, with the perturbed \ac{FRW} metric
\begin{equation}
	\dd s^2 = \ml(1 + \frac{2Φ}{c^2}\mr) c^2 \dd t^2 - a(t)^2 \ml(1 - \frac{2Φ}{c^2}\mr) \dd \vec{x}^2,
\end{equation}
where $a$ is the scale factor and $Φ$ corresponds to the Newtonian potential.
This results in the Schrödinger equation
\begin{equation}
	\label{eq:schroedinger-phys}
	\imag ℏ\ml(∂_t ψ + \frac{3}{2} H ψ\mr) = -\frac{ℏ^2}{2m} ∇^2 ψ +
        mΦψ ,
\end{equation}
where $Φ$ obeys the usual Poisson equation.

Since the scalar particles are bosons with low velocities, most of the particles will be in the ground state.
This enables the use of the mean field approximation, where $ψ$ in \cref{eq:schroedinger-phys} is interpreted as the single macroscopic wave function of a Bose–Einstein condensate with mass density
\begin{equation}
	\label{eq:mass-density}
	ρ = m|ψ|^2.
\end{equation}
\Cref{eq:schroedinger-phys} and the Poisson equation can additionally be rewritten using the \q{comoving} quantities
\begin{equation}
	ψ_{\com} = a^{\sfrac{3}{2}} ψ,
	\quad
	∇_{\com} = a∇,
	\quad
	Φ_{\com} = aΦ,
\end{equation}
to yield the non-linear Schrödinger–Poisson system of equations:\footnote{%
	Due to its origin from the mean field approximation, the \q{Schrödinger equation} \cref{eq:fdm-schroedinger} is technically a Gross–Pitaevskii equation, and the system of \cref{eq:fdm-schroedinger,eq:fdm-poisson} is also known as the Gross–Pitaevskii–Poisson system of equations.%
}
\begin{align}
	\label{eq:fdm-schroedinger}
	\imag ℏ ∂_t ψ_{\com}(t, \vec{x}) &= -\frac{ℏ^2}{2ma(t)^2} ∇_{\com}^2 ψ_{\com}(t, \vec{x}) + \frac{m}{a(t)} Φ_{\com} ψ_{\com}(t, \vec{x})
	\\
	\label{eq:fdm-poisson}
	∇_{\com}^2 Φ_{\com}(t, \vec{x}) &= 4πGm \ml(|ψ_{\com}(t, \vec{x})|^2 - ⟨|ψ_{\com}|^2⟩(t)\mr)
\end{align}
The field $ψ$ is now a function of time and space whose values are ordinary complex numbers, instead of the operator-valued field that was the starting point in \cref{eq:action}.
The challenge for numerical simulations is then to solve the non-linear time evolution of these \emph{\ac{FDM} equations}.

The Schrödinger equation has the form of a diffusion equation and thus has a well-known conservation law described by the continuity equation
\begin{gather}
	\label{eq:continuity}
	∂_t ρ_{\com} + ∇_{\com} ⋅ ρ_{\com} \vec{v}_{\com} = 0,
	\intertext{with the density current}
	\label{eq:density-current}
	ρ_{\com} \vec{v}_{\com}
	= \frac{ℏ}{2\imag} (ψ_{\com}^* ∇_{\com} ψ_{\com} - ψ_{\com} ∇_{\com} ψ_{\com}^*)
\end{gather}
($ρ_{\com} = m|ψ_{\com}|^2$).
The (comoving) velocity field $\vec{v}_{\com}$ gives the peculiar velocity of matter at each point.
The interpretation of the wave function becomes clearer when written in polar form,\footnote{%
	The inclusion of the comoving factor $a^{\sfrac{3}{2}}$ has no impact on the wave function's phase $α$, i.\,e.\ $ψ_{\com} = √{ρ_{\com} / m}\, \euler^{\imag α}$ and $ψ = √{ρ / m}\, \euler^{\imag α}$.%
}
\begin{equation}
	\label{eq:wave-function-polar}
	ψ_{\com} = √{\frac{ρ_{\com}}{m}} \euler^{\imag α},
\end{equation}
with absolute value $√{ρ_{\com} / m}$ and phase $α$, where $ρ_{\com} = a^3 ρ = a^3 m|ψ|^2$ is indeed the same mass density as in \cref{eq:mass-density}.
Inserting this into the expression for the current in \cref{eq:density-current} yields
\begin{equation}
	\label{eq:velocity}
	\vec{v}_{\com} = \frac{ℏ}{m} ∇_{\com} α ,
\end{equation}
that is, the velocity field is given by the gradient of the wave function's phase.
The wave function in polar components (\cref{eq:wave-function-polar}) can be rewritten in hydrodynamical form using the \citet{Madelung1927} transformation, which results in the continuity \cref{eq:continuity} along with a modified Euler equation
\begin{equation}
	∂_t \vec{v}_{\com} + \frac{1}{a^2} ∇_{\com} \vec{v}_{\com}^2
	= -\frac{1}{a} ∇_{\com} Φ_{\com} + \frac{1}{a^2} \frac{ℏ^2}{2m^2} ∇_{\com} \frac{∇_{\com}^2 √{ρ_{\com}}}{√{ρ_{\com}}}
\end{equation}
(cf.\ e.\,g.\ \citet{2020MNRAS.494.2027M}).
This allows for a hydrodynamical interpretation of the density and velocity fields $ρ_{\com}$ and $\vec{v}_{\com}$.

It should be noted that \cref{eq:fdm-schroedinger,eq:fdm-poisson} only have a single parameter given by the constant $ℏ / m$,\footnote{%
	This can be made explicit by using the variable $ψ' = √m ψ$.%
} which is related to the de Broglie wavelength as
\begin{equation}
	\label{eq:de-broglie}
	λ_{\mtext{dB}} = \frac{2πℏ}{mv} .
\end{equation}
Another relevant scale, also determined by $ℏ / m$, is the \ac{FDM} Jeans length.
Due to the existence of \q{quantum pressure} resulting from the Heisenberg uncertainty principle, at least in linear theory, there exists a length scale where this \q{pressure} cancels out the gravitational attraction of matter, analogous to ordinary pressure e.\,g.\ for baryons.
Perturbations at length scales larger than the Jeans length grow, while smaller perturbations oscillate.
The comoving Jeans wave number $k_{\mtext{J}}$ at some redshift $z$ is given by \citep{Hu2000}
\begin{equation}
  \label{eq:jeans}
  k_{\mtext{J}} =
  	\ml(\frac{6 Ω_{\mtext{m}}}{1 + z}\mr)^{\sfrac{1}{4}} \ml(\frac{m H_0}{ℏ}\mr)^{\sfrac{1}{2}}.
\end{equation}

\section{Numerical methodology}
\label{sec:numerics}

Our simulations are performed within the same framework as ordinary cosmological \ac{LCDM} simulations.
The simulation volume consists of a cubic box of side length $L$ with periodic boundary conditions, which is intended to sample the matter distribution in the universe.
The volume is filled with matter whose average comoving density is the mean
background density
\begin{equation}
	⟨ρ_{\com}⟩ = Ω_{\mtext{m}} ρ_{\mtext{crit}} = Ω_{\mtext{m}} \frac{3 H_0^2}{8πG}.
\end{equation}
In order to solve \cref{eq:fdm-schroedinger,eq:fdm-poisson}, a 2nd-order symmetrised split-step pseudo-spectral Fourier method (\q{kick–drift–kick}) is employed.
For a small time step $Δt$, the time evolution can be approximated as follows \citep{2009ApJ...697..850W,2018JCAP...10..027E}:
\begin{align}
	&ψ_{\com}(t + Δt, \vec{x})
	\notag
	\\
	&\;
	= 𝒯 \euler^{-\imag ∫_t^{t + Δt} \ml(-\frac{ℏ}{2m} \frac{1}{a(t')^2} ∇_{\com}^2 + \frac{m}{ℏ} \frac{1}{a(t')} Φ_{\com}(t', \vec{x})\mr) \dd t'} ψ_{\com}(t, \vec{x})
	\displaybreak[0]
	\\
	&\;
	≈ \euler^{\imag \frac{Δt}{2} \ml(\frac{ℏ}{m} \frac{1}{a(t)^2} ∇_{\com}^2 - \frac{m}{ℏ} \frac{1}{a(t)} Φ_{\com}(t + Δt, \vec{x}) - \frac{m}{ℏ} \frac{1}{a(t)} Φ_{\com}(t, \vec{x})\mr)} ψ_{\com}(t, \vec{x})
	\displaybreak[0]
	\\
	&\;
	≈ \underbrace{\euler^{-\imag \frac{m}{ℏ} \frac{1}{a(t)} \frac{Δt}{2} Φ_{\com}(t + Δt, \vec{x})}}_{\text{\q{kick}}} \:
	\underbrace{\euler^{\imag \frac{ℏ}{m} \frac{1}{a(t)^2} \frac{Δt}{2} ∇_{\com}^2}}_{\text{\q{drift}}} \:
	\underbrace{\euler^{-\imag \frac{m}{ℏ} \frac{1}{a(t)} \frac{Δt}{2} Φ_{\com}(t, \vec{x})}}_{\text{\q{kick}}} \,
	ψ_{\com}(t, \vec{x})
\end{align}
where $𝒯$ is the time ordering operator and, using the Baker–\allowbreak Campbell–\allowbreak Hausdorff formula, the time evolution operator has been split into three parts which do not mix
functions of the position and derivative operators.

The fields $ψ$ and $Φ$ are discretised on a uniform Cartesian mesh with $N^3$ grid points to allow for numerical computations using the \ac{FFT}.
Accordingly, the numerical algorithm of the pseudo-spectral method performs the following operations:%
\begin{subequations}
\begin{alignat}{3}
	\label[algorithm]{eq:algorithm-kick1}
	&{\bullet}\quad
	&
	ψ_{\com} &← \euler^{-\imag \frac{m}{ℏ} \frac{1}{a} \frac{Δt}{2} Φ_{\com}} ψ_{\com}
	&
	\text{(kick)}&
	\\
	\label[algorithm]{eq:algorithm-drift}
	&{\bullet}
	&
	ψ_{\com} &← \FFT^{-1}\ml(\euler^{-\imag \frac{ℏ}{m} \frac{1}{a^2} \frac{Δt}{2} k^2} \FFT(ψ_{\com})\mr)
	\hspace*{-8em}&
	\text{(drift)}&
	\\
	&{\bullet}
	&
	Φ_{\com} &← \FFT^{-1}\ml(
		-\frac{1}{k^2} \FFT\ml(4πGm\ml(|ψ_{\com}|^2 - ⟨|ψ_{\com}|^2⟩\mr)\mr)
	\mr)
	\hspace*{-8em}&
	&
	\notag
	\\
	\label[algorithm]{eq:algorithm-poisson}
	&&&&
	\text{(update potential)}&
	\\
	\label[algorithm]{eq:algorithm-kick2}
	&{\bullet}
	&
	ψ_{\com} &← \euler^{-\imag \frac{m}{ℏ} \frac{1}{a} \frac{Δt}{2} Φ_{\com}} ψ_{\com}
	&
	\text{(kick)}&
	\\
	\label[algorithm]{eq:algorithm-last}
	&{\bullet}\quad
	\mathrlap{\text{Go to \cref{eq:algorithm-kick1}}}
\end{alignat}
\end{subequations}
Consecutive executions of \cref{eq:algorithm-kick1,eq:algorithm-kick2} (i.\,e.\ except for the initial and final time steps) can be combined into a single operation $ψ_{\com} ← \euler^{-\imag \frac{m}{ℏ} Δt Φ_{\com}} ψ_{\com}$ to improve performance.

The choice of the time step $Δt$ in \crefrange{eq:algorithm-kick1}{eq:algorithm-last} is determined by the requirement that the phase difference in the exponentials must not exceed $2π$, at which point the time step would be incorrectly \q{aliased} to a smaller time step corresponding to the phase difference subtracted by a multiple of $2π$ due to the periodicity of the exponential function.
Both the kicks (\cref{eq:algorithm-kick1,eq:algorithm-kick2}) and the drift (\cref{eq:algorithm-drift}) yield separate constraints for $Δt$, which must be simultaneously fulfilled.
The resulting time step criterion is
\begin{equation}
	\label{eq:time-step}
	Δt < \min\ml(
		\frac{4}{3π} \frac{m}{ℏ} a^2 Δx^2,\;
		2π \frac{ℏ}{m} a \frac{1}{|Φ_{\com}|_{\mtext{max}}}
	\mr),
\end{equation}
where $Δx = L / N$ is the spatial resolution and $|Φ_{\com}|_{\mtext{max}}$ is the maximum (absolute) value of the potential.
The constraint involving the resolution $Δx$ stems from the drift operation, while the constraint with the potential $Φ_{\com, \mtext{max}}$ results from the kick operation.
The dependence $Δt ∝ Δx^2$ seems to be typical for all numerical
approaches to the Schrödinger–Poisson system of 
\cref{eq:fdm-schroedinger,eq:fdm-poisson}, and can be viewed as
a reflection of the relation of the Schrödinger equation to 
diffusion problems.

Another constraint on the validity of the discretisation becomes apparent when considering the velocity field (\cref{eq:velocity}).
Since this is given by the \emph{gradient} of the wave function's phase, whose difference between two points can be at most $2π$, it follows that the discretised velocity field cannot exceed a maximum value (depending on the concrete form of the discretised gradient operator) of about
\begin{equation}
	\label{eq:velocity-criterion}
	v_{\mtext{max}} = \frac{ℏ}{m} \frac{π}{Δx}.
\end{equation}
Velocities $v ≥ v_{\mtext{max}}$ cannot be represented in a simulation with resolution $Δx$.
Comparing to \cref{eq:de-broglie}, it becomes apparent that this statement is equivalent to a constraint on the resolution, which should be good enough to resolve the de Broglie wavelength of the largest velocities:
\begin{equation}
	\label{eq:velocity-criterion-resolution}
	Δx < \frac{πℏ}{mv_{\mtext{max}}} = \frac{1}{2} λ_{\mtext{dB}}(v_{\mtext{max}}).
\end{equation}

The requirement to resolve structures on the scale of the de Broglie wavelength, given by \cref{eq:velocity-criterion-resolution}, in combination with the time step requirement $Δt ∝ Δx^2$ (\cref{eq:time-step}), is one of the main reasons why \ac{FDM} simulations require so many more computational resources than traditional particle-based \ac{CDM} simulations.

To enable our simulation work, the pseudo-spectral algorithm
(\crefrange{eq:algorithm-kick1}{eq:algorithm-last}) has been
implemented as a module in the {\AREPO} code
\citep{2010MNRAS.401..791S,Weinberger2020} in a similar approach to
\citet{2017MNRAS.471.4559M,2020MNRAS.494.2027M}. \Cref{eq:algorithm-poisson} is
performed using \AREPO's existing Poisson solver algorithm, while the
split-step solution to the Schrödinger equation is solved using newly
developed, highly parallel code inspired by the same algorithm. It,
too, makes use of the \q{\ac{FFTW}} library \citep{Frigo2005} to perform
the \ac{FFT}.

This new implementation enables simulations of meshes
which are several orders of magnitude larger than any previous work.
The split-step scheme and the integration with {\AREPO} also make it
suitable for future simulations including baryons.
All \ac{FDM} simulations reported in this work were performed using
this pseudo-spectral \ac{FDM} {\AREPO} module, whereas
the \ac{CDM} simulations were done using \AREPO's standard TreePM method.

\begin{figure*}
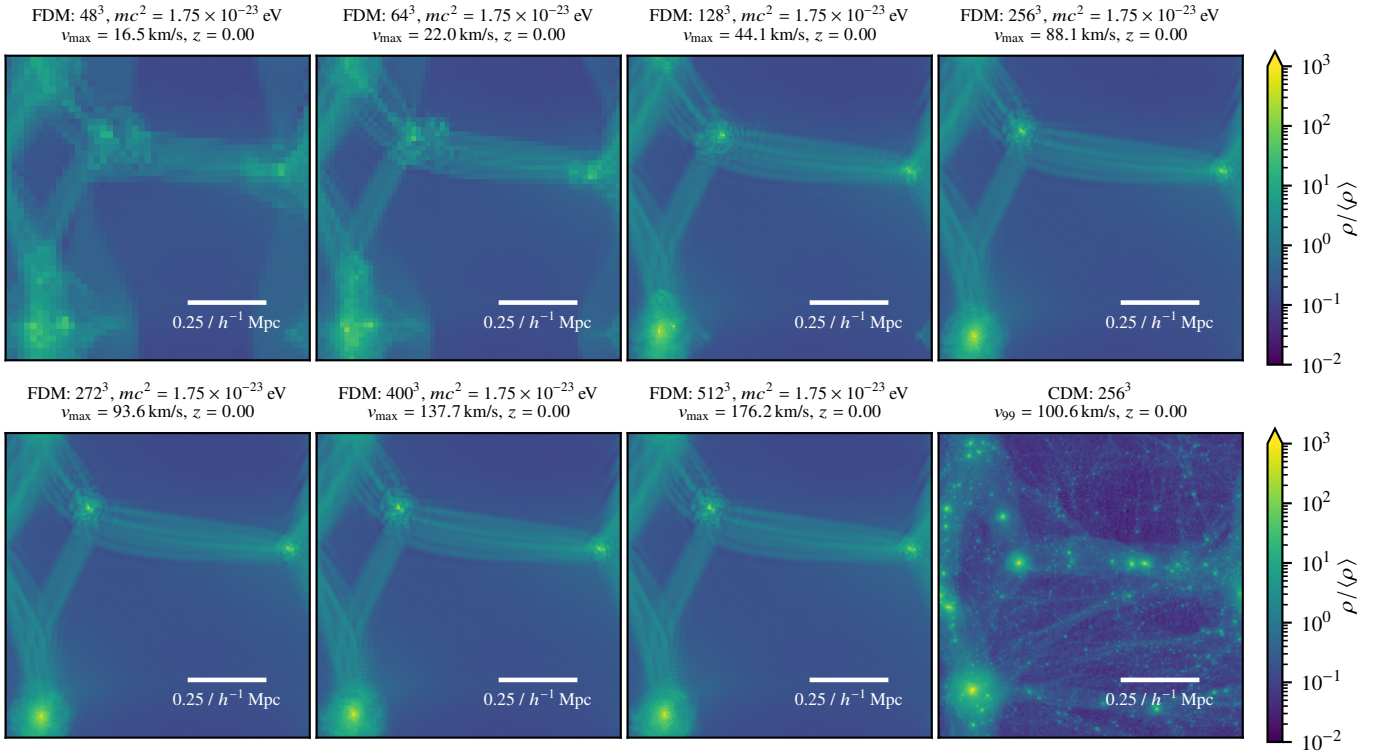

	\centering
	\import{res/1mpc_resolutions/}{row1_grid_project.pgf}

	\smallskip

	\import{res/1mpc_resolutions/}{row2_grid_project.pgf}

	\caption{%
		Projected dark matter density at $z = 0$ in $L = \SI{1}{\per\hHubble\Mpc}$ cosmological box simulations of \acs*{FDM} with $m c^2 = \SI{1.75e-23}{\eV}$ and \acs*{CDM} \acsp*{IC} for different resolutions.
		A high-resolution \acs*{CDM} simulation is shown for comparison.
		The 99th-percentile velocity of the particles in the \acs*{CDM} simulation is given as $v_{99}$.%
	}
	\label{fig:1mpc-projections}
\end{figure*}

\subsection{Initial conditions}
\label{sec:ics}

As mentioned previously, there is in fact a correspondence between the Vlasov–Poisson equations which describe the evolution of a \ac{CDM} fluid and the Schrödinger–Poisson equations, with the latter reducing to the former for $ℏ / m → 0$ \citep{1993ApJ...416L..71W,2018PhRvD..97h3519M,2020JCAP...04..003G}.
A wave function $ψ(\vec{x})$ can be constructed from a phase space
distribution function $f(\vec{x}, \vec{p})$ by means of a Wigner or Husimi transform.
This relationship illustrates why \ac{FDM} behaves like \ac{CDM} on scales larger than $λ_{\mtext{dB}}$.
At first, this was in fact considered as an alternate method to simulate \ac{CDM} \citep{1993ApJ...416L..71W}.
However, one can also take the equations seriously and actually consider the Schrödinger–Poisson equations as the \q{true} description of dark matter.
In this case, the correspondence allows one to \q{translate} a phase space distribution function to the wave function formalism.
This is very useful because it enables direct comparisons between structure formation with \ac{CDM} and \ac{FDM} from the same \acp{IC}.

In general, a wave function on a discretised lattice can be constructed using the prescription
\begin{equation}
	ψ(\vec{x}_{\vec{n}})
	∝ ∑_{n_1'=0}^{N - 1} … ∑_{n_3'=0}^{N - 1}
		√{f(\vec{x}_{\vec{n}'}, \vec{v}_{\vec{n}})} \,
		\euler^{\imag\frac{m}{ℏ} \vec{x}_{\vec{n}'} ⋅ \vec{v}_{\vec{n}} + R_{\vec{n}}},
\end{equation}
where $\vec{n}$, $\vec{n}' = (n_1', n_2', n_3')$ are discrete grid
indices, $\vec{x}_{\vec{n}} = \vec{n} Δx$ and $\vec{v}_{\vec{n}} =
2π\frac{ℏ}{m} \frac{\vec{n}}{N Δx}$ are discrete phase space grid
points, and $R_{\vec{n}}$ is a random phase which is required to
ensure that different velocity components are uncorrelated \citep[cf.][]{1993ApJ...416L..71W,2018PhRvD..97h3519M}.

For the case of a \q{cold} or \q{single-stream} distribution function
as in the case of \ac{CDM}, where each point in space has a single
well-defined value for the velocity, this value can be directly related to the wave function's phase as in \cref{eq:velocity}.
The construction of the wave function then simply reduces to the polar decomposition (\cref{eq:wave-function-polar}), with the absolute value and phase determined by the density and velocity at each point, respectively:
\begin{gather}
	\label{eq:ic-absolute}
	|ψ(\vec{x})| = √{\frac{ρ(\vec{x})}{m}},
	\displaybreak[0]
	\\
	\label{eq:ic-phase}
	∇ \arg\ml(ψ(\vec{x})\mr)
	= ∇ α(\vec{x})
	= \frac{m}{ℏ} \vec{v}(\vec{x}).
\end{gather}
\Cref{eq:ic-phase} can be easily solved numerically by applying the spectral method to the equation
\begin{equation}
	∇^2 α(\vec{x})
	= \frac{m}{ℏ} ∇ ⋅ \vec{v}(\vec{x}).
\end{equation}

The \acp{IC} were generated using the {\NGenIC} code \citep{2015ascl.soft02003S}, which employs the Zel'dovich approximation to generate a perturbed particle distribution, for an ordinary \ac{LCDM} cosmological simulation at the starting redshift $z = 1/a - 1 = 127$ with an input power spectrum following \citet{Efstathiou1990,Efstathiou1992}, i.\,e.\ of the form
\begin{equation}
	P(k) ∝
	k \ml(1 + \ml(ak + (bk)^{\frac{3}{2}} + c^2 k^2\mr)^ν\mr)^{-\frac{2}{ν}}
\end{equation}
with $a = \num{6.4}/Γ\,\si{\per\hHubble\Mpc}$, $b = \num{3.0}/Γ\,\si{\per\hHubble\Mpc}$, $c = \num{1.7}/Γ\,\si{\per\hHubble\Mpc}$, $Γ = Ω_{\mtext{m}}h = \num{0.21}$,\footnote{%
	See \cref{sec:simulations}.%
} and $ν = \num{1.13}$.
A wave function was constructed from the same \acp{IC} using the procedure in \cref{eq:ic-absolute,eq:ic-phase}.

\subsection{Resolution and convergence tests}

Before using the newly-developed \ac{FDM} {\AREPO} module for simulations of larger cosmological boxes, extensive tests were performed on boxes of comoving size $L = \SI{1}{\per\hHubble\Mpc}$.\footnote{%
	The cosmological parameters are the same as in \cref{sec:simulations}.%
}
This not only allowed for verification of the code's correctness by comparing to \ac{CDM} simulations with the same \acp{IC} and to other implementations of \ac{FDM}, but also to study the behaviour of the pseudo-spectral method.
Of particular interest are the resolution requirements, e.\,g.\ to
what extent the numerical results remain valid and the convergence of
the matter power spectrum is compromised when the velocity constraint (\cref{eq:velocity-criterion}) is violated.
Moreover, the differences to the behaviour of \ac{CDM} simulations are of interest even in this small test volume.
Although the simulation volume is too small to be cosmologically representative at $z = 0$ since even the largest scales become non-linear by then, there are opportunities to observe what differences (or similarities) become apparent in both the linear (early times) and non-linear (late times) regimes of both dark matter models.

\Cref{fig:1mpc-projections} shows the projected density at $z = 0$ of
such test simulations across a wide range of resolutions, with a \ac{CDM} simulation for comparison.\footnote{%
	Due to the fact that at $z = 0$, all scales have become non-linear with such a small box size, the largest-scale structures parallel to the coordinate axes (\q{fundamental modes}, i.\,e.\ the modes with the minimal value of $k = 2π/L$) are clearly visible.%
}
Notably, even at the lowest resolution, the qualitative features are preserved and structures on the largest scales remain the same.
However, for the lower resolutions, structures become increasingly \q{smeared out} (filling a larger volume), and the resulting range of values in the density contrast is decreased, indicating that the lack of resolution interferes with the formation of more compact structures.
On the other hand, for the higher resolutions, there are no appreciable discrepancies in the low-density regions and filaments, with the only differences being slight changes in the position and internal structure of the largest haloes.
This is in line with the idea of the velocity criterion \cref{eq:velocity-criterion} and indicates that simulations can yield valid answers in regions with low velocities even in the presence of high-velocity regions which violate the criterion.
Common to all the \ac{FDM} simulations, and in contrast to the \ac{CDM} simulation, almost all small-scale structure is erased due to the Heisenberg uncertainty principle with a large de Broglie wavelength ($λ_{\mtext{dB}} = \SI{1.21}{\kpc}$ for $m c^2 = \SI{e-22}{\eV}$, $v = \SI{100}{\km\per\s}$).
The \ac{FDM} structures consist of a few massive haloes and smooth overdense filaments, while in \ac{CDM}, the filaments fragment into sub-haloes down to the smallest scales.

\begin{figure}
	\centering
	\import{res/}{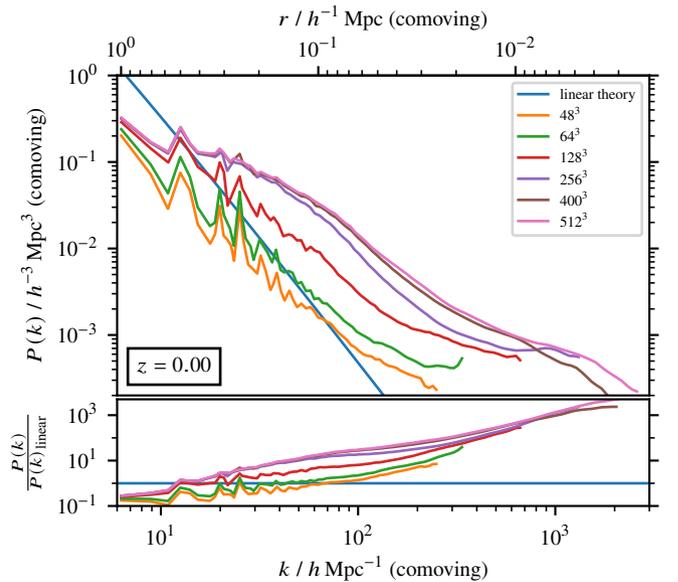}

	\caption{%
		Dark matter power spectra for cosmological \acs*{FDM} simulations with different resolutions, box size $L = \SI{1}{\per\hHubble\Mpc}$, and \acs*{FDM} mass $m c^2 = \SI{1.75e-23}{\eV}$, at $z = 0$ evolved from \acs*{CDM} \acsp*{IC}.
		The legend indicates the grid size for each included simulation.
		The power spectrum evolved using linear perturbation theory is shown for comparison.
		The bottom panel shows the ratio of the power spectra to the result from linear theory.%
	}
	\label{fig:1mpc-powerspec}
\end{figure}

A more quantitative indication of numerical convergence is given in \cref{fig:1mpc-powerspec} by the power spectra of the density field.
This demonstrates that, for the given setup, grid sizes of $128^3$ and smaller lead to significant deviations from higher-resolution results even on large scales.
This discrepancy is visible even in the density projections of \cref{fig:1mpc-projections}.
On the other hand, while it becomes difficult to see any visual differences for grid sizes of $256^3$ and larger, the power spectra show that, to some degree, even the $256^3$ grid suffers from lack of power on smaller scales.
In this case, at least roughly, the velocity criterion seems to give a rather good indication of the resolution required to achieve convergence of the power spectrum on most scales (cf.\ the values of $v_{\mtext{max}}$ in \cref{fig:1mpc-projections}).
Generally, the (relative) lack of power is most pronounced and persists for higher resolutions on smaller scales, i.\,e.\ the power spectrum converges progressively from larger to smaller scales with increasing resolution, although even convergence on just the largest scales places significant demands on resolution (better than a $128^3$ grid in this case).

\subsection{Computational cost analysis}

As mentioned before, the required computational resources – both run-time cost and memory – are a significant obstacle when attempting to perform large-scale simulations of \ac{FDM}.
The run-time cost depends on three simulation parameters: the \ac{FDM} mass $m$, the (comoving) box size $L$ and the number of mesh points $N^3$ (where $N$ is the number of points per dimension).\footnote{%
	Instead of $N$, the resolution $Δx = L/N$ can be equivalently considered.%
}

\begin{figure}
	\centering
	\import{res/}{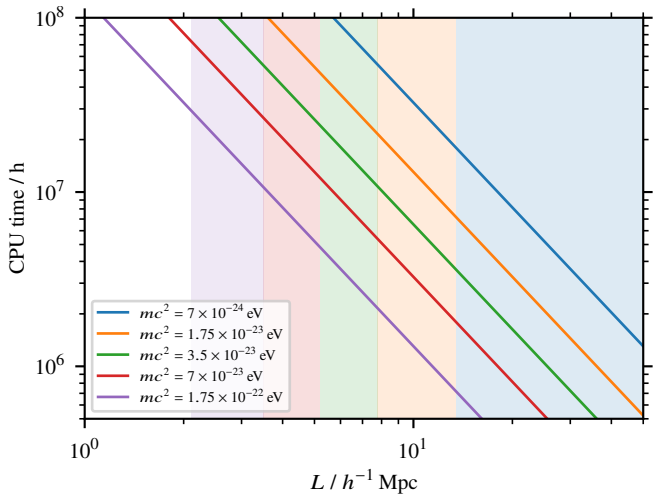}

	\caption{%
		The expected computational cost of cosmological simulations when running to $z = 0$ as a function of box size $L$ for a mesh size of $N^3 = 8192^3$ and different \acs*{FDM} masses.
		Shaded areas indicate regions where the velocity constraint \cref{eq:velocity-criterion} is violated for the corresponding mass at $z = 0$ due to lack of resolution.
		The concrete values of CPU time correspond to the Cobra cluster at \acs*{MPCDF} for reference.%
	}
	\label{fig:sim-cost-L}
\end{figure}

\begin{figure}
	\centering
	\import{res/}{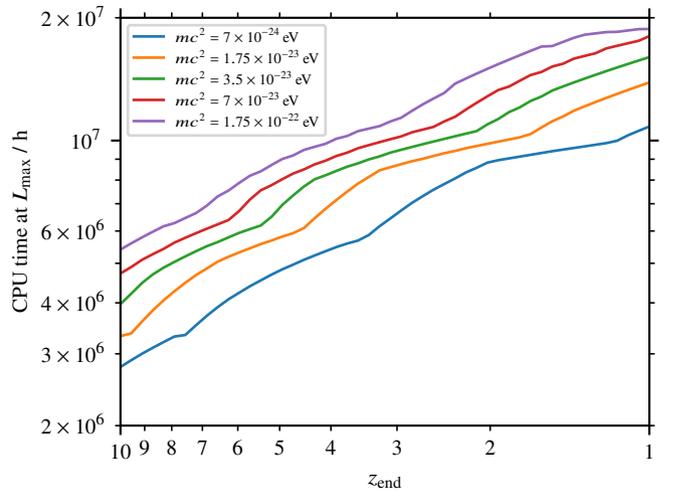}

	\caption{%
		The expected computational cost of cosmological simulations with different \acs*{FDM} masses as a function of target redshift $z_{\mtext{end}}$ for a mesh size of $N^3 = 8192^3$ at the maximum allowed box size $L = L_{\mtext{max}}$, as given by the velocity constraint \cref{eq:velocity-criterion}.
		The points where the curves intersect their associated shaded regions in \cref{fig:sim-cost-L} would correspond to $z_{\mtext{end}} = 0$ in this figure.%
	}
	\label{fig:sim-cost-z}
\end{figure}

For each individual time step, the operations in \crefrange{eq:algorithm-kick1}{eq:algorithm-last} are performed.
The cost of a single time step only depends on the mesh size $N^3$, while $m$ and $L$ only enter the computation through multiplication as constant factors.
It is dominated by the cost of the \ac{FFT} operations, which scales as $O\big(N^3 \ln(N^3)\big)$.
The total cost then depends on the number of time steps which have to be performed, which is determined by the time step size given by \cref{eq:time-step}.
The drift criterion $∝ Δx^2$ is typically more stringent, especially in light of the resolution requirements (e.\,g.\ \cref{eq:velocity-criterion}) and for higher redshifts.
Thus, the total number of time steps is roughly inversely proportional to the limit given by the drift criterion, which results in the following approximate behaviour of the computation time $C$:
\begin{equation}
	\label{eq:cpu-time}
	C ∝ \frac{N^5 \ln(N^3)}{m L^2}.
\end{equation}
This relation was also verified in practice using measurements of small test computations.

Particularly striking is the strong dependence of \cref{eq:cpu-time}
on $N^5$, which rapidly leads to exploding computational costs for
larger $N$.
At fixed $N$, the cost can be decreased by choosing a larger box size $L$ because this reduces the resolution $Δx$.
However, as mentioned before, the resolution requirements are quite strict due to the inherent scale in the Schrödinger–Poisson system, which must be resolved in order to yield reliable results.

To give a tangible reference for the absolute costs involved on the one hand, but also the code's performance on the other hand, \cref{fig:sim-cost-L} shows the expected cost of simulations with mesh size $N^3 = 8192^3$ to $z = 0$ in CPU hours on the state-of-the-art Cobra supercomputer at the \ac{MPCDF}.\footnote{%
	\url{https://www.mpcdf.mpg.de/services/supercomputing/cobra}%
}
\Cref{fig:sim-cost-L} displays the CPU time for a range of masses $m$ with variable box size $L$.
Even when allowing for the velocity criterion \cref{eq:velocity-criterion} to be violated, it becomes clear that such a simulation would take at least \SI{e7}{\CPU\hour}.
Even worse, the resolution requirements severely limit the possible simulation volume to $L < \SI{10}{\per\hHubble\Mpc}$ for the mass range of interest.
Since the de Broglie wavelength increases with smaller masses, the resolution requirements can be loosened by choosing much smaller values for $m$, but this does not nearly go far enough as the physically viable range is bounded around $≳ \SI{e-22}{\eV}$.
The computational cost for the largest possible box size $L_{\mtext{max}}$ at a given mass $m$ (intersections of the graphs with the vertical lines) decreases only weakly with decreasing $m$.

\Cref{fig:sim-cost-z} shows the same computation time at the largest possible box size due to \cref{eq:velocity-criterion} for each mass, but now for the case where the simulation is stopped earlier than $z = 0$.\footnote{%
	The unevenness in the lines results due to \AREPO's time binning procedure, which \q{discretises} the allowed time step values.%
}
Unfortunately, the computational cost does not depend very strongly on
the final redshift (approximately $∝ z^{-\sfrac{1}{2}}$ in the region shown); it is merely lowered by a factor of $≲ 5$ even when stopping at $z = 6$, and still remains at several million CPU hours for a $N^3 = 8192^3$ simulation.

\begin{table}
	\centering
	\caption{%
		List of performed simulations with important characteristics.
		The lengths given for the box sizes and resolutions are comoving.%
	}
	{%
	\sisetup{list-pair-separator={, }, list-final-separator={, }}%
	\setlength{\tabcolsep}{4.8pt}%
	\begin{tabular}{
		l r S
		c
		S
	}
		\toprule
		Type &
		Res.\ el. &
		{$L$ / \si{\per\hHubble\Mpc}} &
		{$m c^2$ / \si{\eV}} &
		{Resolution}
		\\
		\midrule
		FDM & $8640^3$ & 10 & \hfill \num{7e-23} & \SI{1.16}{\per\hHubble\kpc}
		\\
		FDM & $4320^3$ & 10 & (\numlist{3.5; 7}) $×$ \num{e-23} & \SI{2.31}{\per\hHubble\kpc}
		\\
		FDM & $3072^3$ & 10 & (\numlist{3.5; 7}) $×$ \num{e-23} & \SI{3.26}{\per\hHubble\kpc}
		\\
		FDM & $2048^3$ & 10 & (\numlist{3.5; 7}) $×$ \num{e-23} & \SI{4.88}{\per\hHubble\kpc}
		\\
		FDM & $4320^3$ & 5 & \hfill \num{7e-23} & \SI{1.16}{\per\hHubble\kpc}
		\\
		FDM & $3072^3$ & 5 & (\numlist{3.5; 7}) $×$ \num{e-23} & \SI{1.63}{\per\hHubble\kpc}
		\\
		FDM & $2048^3$ & 5 & (\numlist{3.5; 7}) $×$ \num{e-23} & \SI{2.44}{\per\hHubble\kpc}
		\\
		FDM & $1024^3$ & 5 & (\numlist{3.5; 7}) $×$ \num{e-23} & \SI{4.88}{\per\hHubble\kpc}
		\\[\smallskipamount]
		CDM & $2048^3$ & 10 & {—} & \SI{9.69e3}{\per\hHubble\Msunit}
		\\
		CDM & $1024^3$ & 10 & {—} & \SI{7.75e4}{\per\hHubble\Msunit}
		\\
		CDM & $512^3$ & 10 & {—} & \SI{6.20e5}{\per\hHubble\Msunit}
		\\
		CDM & $1024^3$ & 5 & {—} & \SI{9.69e3}{\per\hHubble\Msunit}
		\\
		CDM & $512^3$ & 5 & {—} & \SI{7.75e4}{\per\hHubble\Msunit}
		\\
		\bottomrule
	\end{tabular}%
	}
	\label{tab:simulations}
\end{table}

\subsection{Simulations}
\label{sec:simulations}

All our simulations were performed using the cosmological parameters $Ω_{\mtext{m}} = \num{0.3}$, $Ω_{\mtext{b}} = \num{0}$, $Ω_Λ = \num{0.7}$, $H_0 = \SI{70}{\km\per\s\per\Mpc}$ ($h = \num{0.7}$), and $σ_8 = \num{0.9}$,\footnote{%
	As usual, the density parameters $Ω_i$ indicate the values at $z = 0$.%
}
with \acp{IC} as described in \cref{sec:ics}.
The same random seed was shared for the generation of all \acp{IC} at $z = 127$ in order to allow for a direct comparison between different dark matter models and resolutions.
In order to avoid the onset of resolution effects that would affect even our highest-resolution simulations, the simulations were run until $z = 3$.
For comoving box sizes of \SIlist{5; 10}{\per\hHubble\Mpc}, and masses $m c^2$ of \SIlist{3.5e-23; 7e-23}{\eV}, simulations with different resolutions – up to mesh sizes of $N^3 = 8640^3$ – were performed.
A detailed list of the different simulations is given in \cref{tab:simulations}.

To date, these are the largest three-dimensional cosmological simulations of structure formation including the full \ac{FDM} dynamics to low redshifts, with a simulation volume which is at least $64$ times larger and a mesh which has at least $600$ times more resolution elements than any comparable existing work.
The largest simulation required more than \SI{7e6}{\CPU\hour} on the Cobra cluster to complete.
Previous efforts (using various methods) have only reached box sizes of \SI{2.5}{\per\hHubble\Mpc} \citep[][using an \ac{AMR} method, until $z = \num{2.5}$]{2020arXiv200704119M} or \SI{1.7}{\per\hHubble\Mpc} \citep[][also with a pseudo-spectral method, until $z = \num{5.5}$]{2020MNRAS.494.2027M} for full \ac{FDM} simulations, and \SI{2.5}{\per\hHubble\Mpc} with a hybrid method \citep[i.\,e.\ not including the \ac{FDM} dynamics everywhere in the simulation volume]{2018PhRvD..98d3509V} for similar cosmological simulations.

\subsection{Grid-based halo finding}
\label{sec:halo-finder}

The \ac{HMF} is an important measure of large-scale structure.
For typical \ac{CDM} simulations, it is determined using an algorithm like \ac{FoF} in {\AREPO}, which identifies haloes by connecting simulation particles (point masses) whose distance to other particles is below a certain threshold.
When using the pseudo-spectral method, however, there are no particles, and the density field is instead represented by a Cartesian mesh.
This means that widely-used algorithms, which operate on the particle distribution, cannot be used to analyse the \ac{FDM} simulations presented here.

\begin{figure}
	\centering
	\import{res/}{fof_cdm_particles_vs_grid_mass_function_1.0000.pgf}
	\caption{%
		The \acs*{HMF} of a cosmological \acs*{LCDM} simulation at $z = 0$ with $L = \SI{10}{\per\hHubble\Mpc}$ and $512^3$ particles, as determined using \AREPO's standard \acs*{FoF} algorithm (\q{particles}) and the newly implemented grid-based halo finder (\q{grid}).
		The latter used the density grid constructed from a \acs*{CiC} mass assignment of the simulation particles onto a mesh with $1024^3$ points as input, with an overdensity threshold of $60$ times the matter background density.
		The bottom panel shows the ratio of the mass functions.
		Note that the masses given here are the \emph{masses of the \acs*{FoF} groups} (i.\,e.\ the sum of all particle or grid cell masses in the group) to give a direct comparison of the two approaches.%
	}
	\label{fig:fof-particles-vs-grid}
\end{figure}

Because of this, a modified version of the \ac{FoF} algorithm was developed in this work to enable the determination of the \ac{HMF} for a discretised density mesh.
Instead of a linking length, this grid-based halo finder uses a density threshold as a parameter.
Adjacent cells in the mesh are linked if their density exceeds the density threshold.
The grid-based halo finder is part of the pseudo-spectral \ac{FDM} {\AREPO} module.

\Cref{fig:fof-particles-vs-grid} demonstrates that this new halo
finder performs well in comparison to the standard particle-based
\ac{FoF} algorithm. Using a \ac{CDM} simulation with $512^3$
particles, the density was represented both using the original
particle data and a Cartesian mesh with $1024^3$ grid points whose
density values were determined using \ac{CiC} mass assignment. The
\ac{HMF} was determined with the particle and grid data as input for
\ac{FoF} and the new grid-based halo finder, respectively. As evident
in \cref{fig:fof-particles-vs-grid}, both procedures show excellent
agreement. Differences arise in the lightest haloes, which are limited
by resolution since density is always spread out over at least one
cell volume, and slight variations are present for the most massive
haloes, where statistical effects play a role since there are only a
few haloes per mass bin.

\begin{figure*}
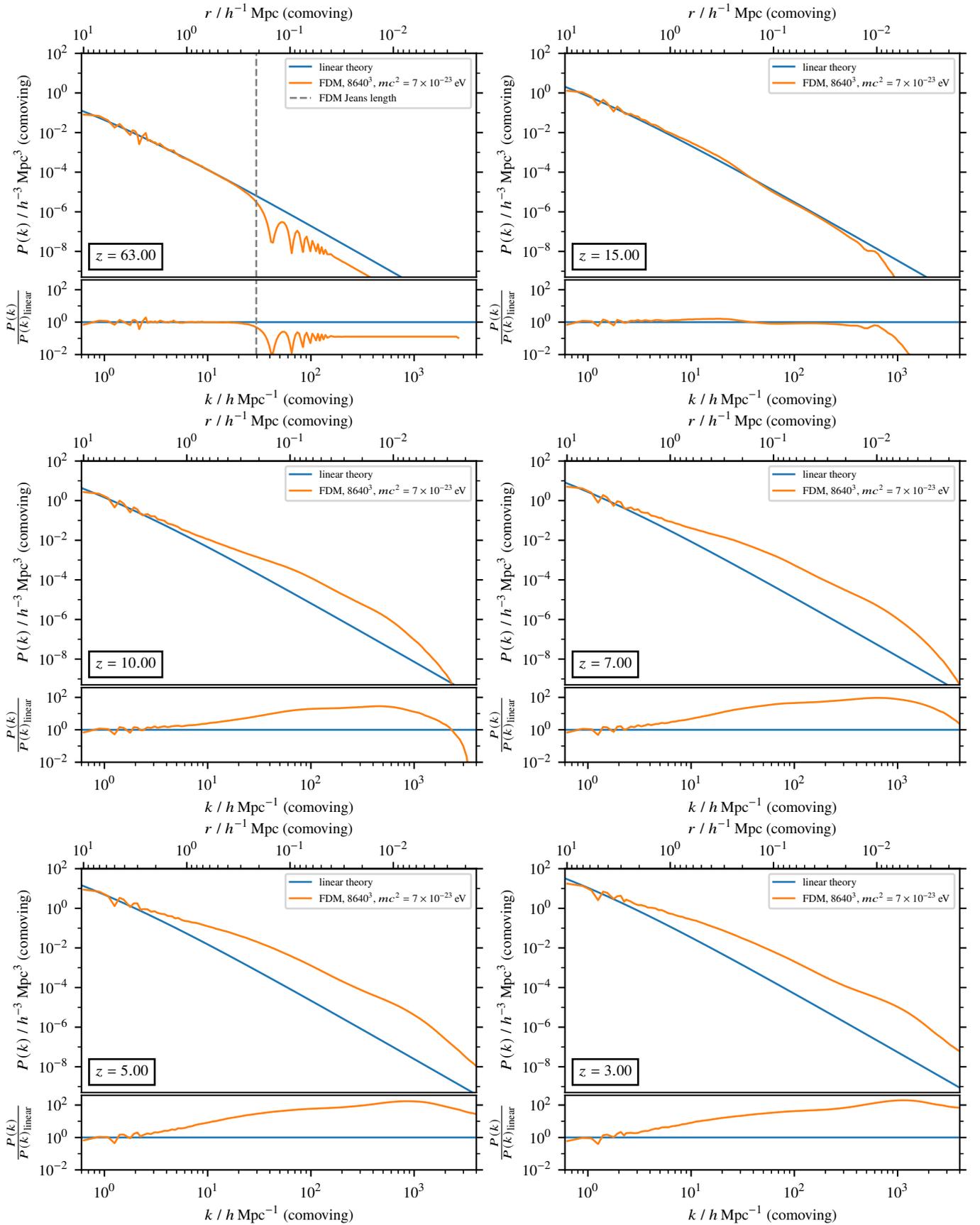

	\centering
	\import{res/psfig1/}{powerspec_0.0156.pgf}%
	\hfill%
	\hspace*{-0.4in}%
	\import{res/psfig1/}{powerspec_0.0625.pgf}

	\vspace*{-10ex}

	\import{res/psfig1/}{powerspec_0.0909.pgf}%
	\hfill%
	\hspace*{-0.4in}%
	\import{res/psfig1/}{powerspec_0.1250.pgf}

	\vspace*{-10ex}

	\import{res/psfig1/}{powerspec_0.1667.pgf}%
	\hfill%
	\hspace*{-0.4in}%
	\import{res/psfig1/}{powerspec_0.2500.pgf}

	\caption{%
		Dark matter power spectra at different redshifts for a high-resolution cosmological \acs*{FDM} simulation with box size $L = \SI{10}{\per\hHubble\Mpc}$, \acs*{FDM} mass $m c^2 = \SI{7e-23}{\eV}$, mesh size $N^3 = 8640^3$, and \acs*{CDM} \acsp*{IC}.
		The power spectrum evolved using linear perturbation theory is shown for comparison.
		The lower panels show the ratio of the power spectra to the result from linear theory.
		For $z = 63$, the dashed line additionally indicates the \acs*{FDM} Jeans scale (\cref{eq:jeans}).%
	}
	\label{fig:power-spectra-fdm-vs-linear}
\end{figure*}

\begin{figure}
	\centering
	\import{res/psfig2/}{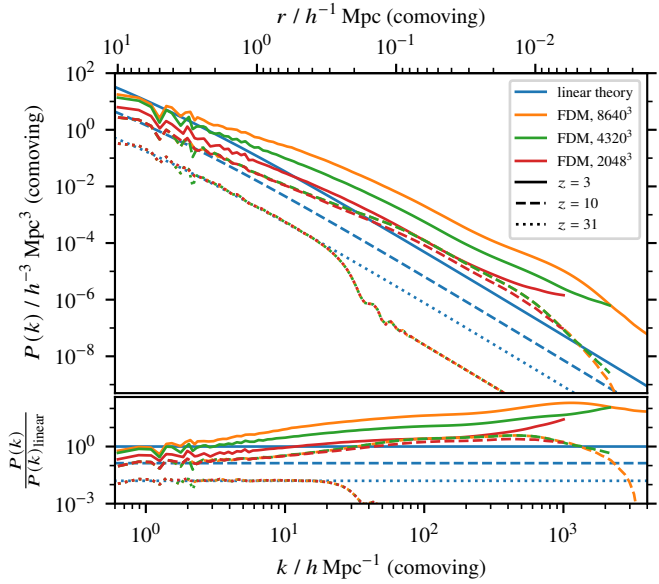}

	\caption{%
		Dark matter power spectra for cosmological \acs*{FDM} simulations with box size $L = \SI{10}{\per\hHubble\Mpc}$, \acs*{FDM} mass $m c^2 = \SI{7e-23}{\eV}$ and \acs*{CDM} \acsp*{IC} at different resolutions (indicated by different colours) and different redshifts (indicated by solid, dashed, and dotted lines).
		The power spectrum evolved using linear perturbation theory is shown for comparison.
		The  bottom panel shows the ratio of the power spectra to the result from linear theory at $z = 3$.%
	}
	\label{fig:power-spectra-resolutions-10mpc}
\end{figure}

\begin{figure}
	\centering

	\import{res/psfig3/}{boxsizes_powerspec_0.2500.pgf}

	\caption{%
		Dark matter power spectra for cosmological \acs*{FDM} simulations at fixed redshift ($z = 3$) with \acs*{FDM} mass $m c^2 = \SI{7e-23}{\eV}$ using \acs*{CDM} \acsp*{IC}, with varying box sizes ($L = \SIlist{5; 10}{\per\hHubble\Mpc}$) and resolutions.
		The power spectrum evolved using linear perturbation theory is shown for comparison.
		The bottom panel shows the ratio of the power spectra to the result from linear theory.%
	}
	\label{fig:power-spectra-boxsizes}

	\import{res/psfig3/}{masses_powerspec_0.2500.pgf}

	\caption{%
		Dark matter power spectra for cosmological \acs*{FDM} simulations at fixed redshift ($z = 3$) with box size $L = \SI{10}{\per\hHubble\Mpc}$  using \acs*{CDM} \acsp*{IC}, with varying \acs*{FDM} masses ($m c^2 = \SIlist{3.5e-23; 7e-23}{\eV}$) and resolutions.
		The power spectrum evolved using linear perturbation theory is shown for comparison.
		The bottom panel shows the ratio of the power spectra to the result from linear theory.%
	}
	\label{fig:power-spectra-masses}
\end{figure}

\begin{figure*}
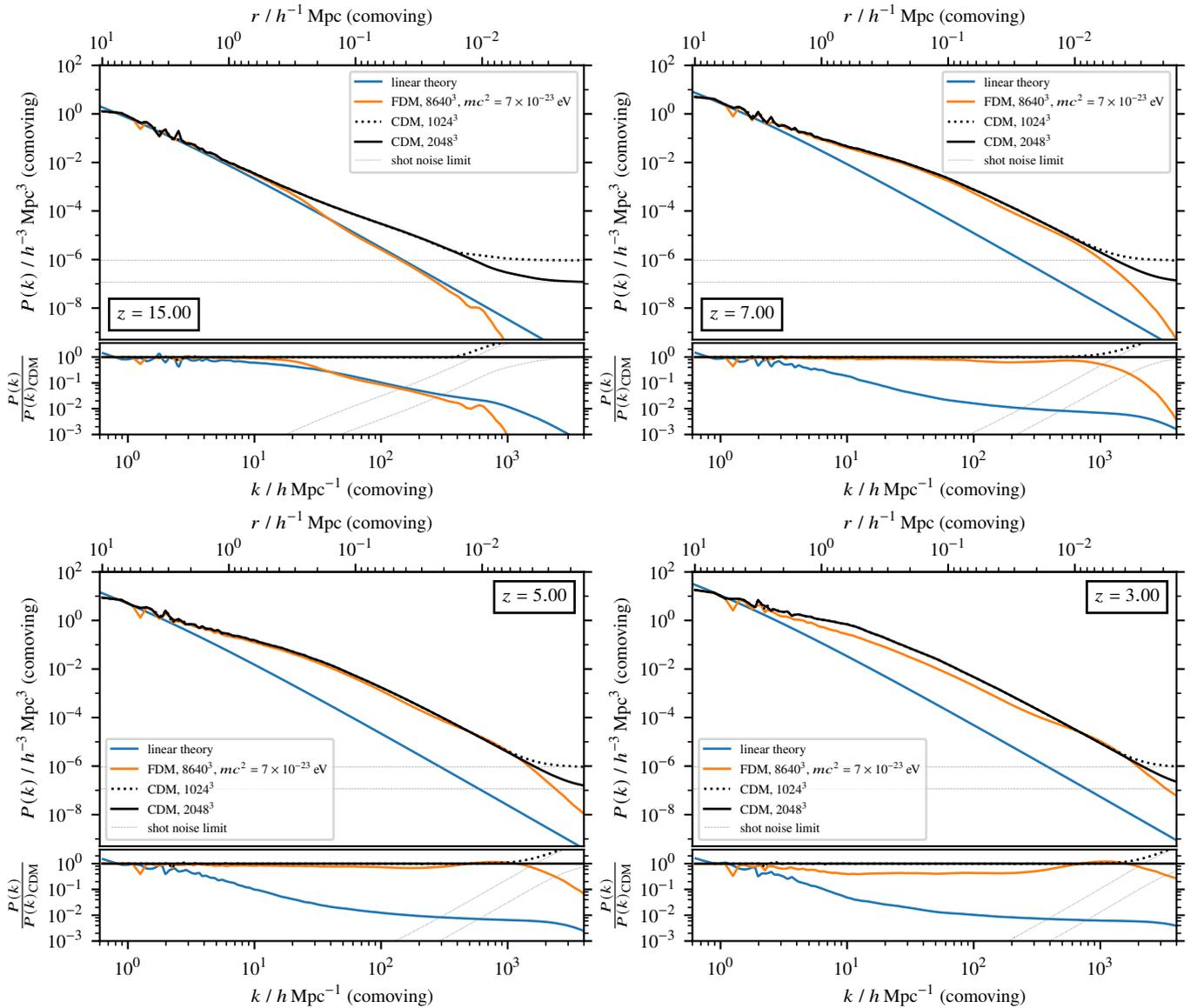

	\centering
	\import{res/psfig4/}{powerspec_0.0625.pgf}%
	\hfill%
	\hspace*{-0.4in}%
	\import{res/psfig4/}{powerspec_0.1250.pgf}

	\vspace*{-10ex}

	\import{res/psfig4/}{powerspec_0.1667.pgf}%
	\hfill%
	\hspace*{-0.4in}%
	\import{res/psfig4/}{powerspec_0.2500.pgf}

	\caption{%
		Dark matter power spectra at different redshifts for a
        high-resolution cosmological \acs*{FDM} simulation
        with box size $L = \SI{10}{\per\hHubble\Mpc}$,
        \acs*{FDM} mass $m c^2 = \SI{7e-23}{\eV}$, mesh size $N^3 = 8640^3$, and
        \acs*{CDM} \acsp*{IC}, compared to cosmological $N$-body
        \acs*{CDM} simulations with different resolutions.
        The power spectrum evolved using linear perturbation theory is shown for comparison.
      	The lower panels show the ratio of the power spectra to the result of the highest-resolution \acs*{CDM} simulation.
      	For $z = 63$, the dashed line additionally indicates the \acs*{FDM} Jeans scale (\cref{eq:jeans}).
      	Faint dotted lines show the shot noise limits of the $N$-body simulations; the power spectrum cannot be measured accurately once it reaches this limit.%
	}
	\label{fig:power-spectra-fdm-vs-cdm}
\end{figure*}

\section{Dark matter power spectrum}
\label{sec:power-spectrum}

The matter power spectrum gives important insight into how matter
clusters at different length scales, and its evolution in time is by
now quite accurately known for \acs{LCDM} \citep[e.\,g.][]{Jenkins1998},
even when baryonic effects are included
\citep{Springel2018}. The power spectrum is also a
particularly important diagnostic for a comparison of \ac{FDM} with
\ac{CDM}, because the former is expected to suppress structure
formation at scales smaller than $λ_{\mtext{dB}}$ due to the
Heisenberg uncertainty principle.
Due to the use of \ac{CDM} \acp{IC}, our results demonstrate to what extent the dynamics of the Schrödinger–Poisson \cref{eq:fdm-schroedinger,eq:fdm-poisson} affect the clustering of matter compared to \ac{CDM}, even without the initial suppression present in a \q{realistic} \ac{FDM} cosmology (cf.~\cref{sec:fdm-ics}).

In \cref{fig:power-spectra-fdm-vs-linear}, we show the time evolution
of the matter power spectra of \ac{FDM} at our highest resolution of
$8640^3$ cells in a \SI{10}{\per\hHubble\Mpc} box. From the top left to
the bottom right panel, we show redshifts from $z = 63$ to $z = 3$, and
compare to expectations from linear theory. In particular, the bottom
of each panel gives the ratio of the measured \ac{FDM} power to the
linear theory power spectrum used for initializing the simulation. At
high redshifts and at small $k$, \ac{FDM} accurately follows linear
theory, with modes growing independently with the same linear growth
factor, just like \ac{CDM} does, so that the pattern of random
fluctuations of our particular realization around the smooth initial
input spectrum is preserved in time. Beginning at scales around
$k ≈ \SI{10}{\hHubble\per\Mpc}$ and a redshift of $z ≈ 15$, signs of mildly
non-linear evolution are apparent, which manifest themselves in a
stronger than linear growth of power. This non-linear evolution
becomes quickly more pronounced in a way reminiscent of \ac{CDM}, except
that on the smallest resolved scales, for
$k ≥ \SI{1000}{\hHubble\per\Mpc}$, the non-linear growth appears sluggish
and lagging behind that seen on larger scales.

It is now important to examine on which scales and at which times
these results are quantitatively reliable. As we have discussed
earlier, the stringent numerical requirements of \ac{FDM} make this
more involved than for \ac{CDM}, because here even the large-scale linear
growth requires a fairly high resolution to get right, and it is not
readily clear how numerical limitations will manifest themselves in
the results. In \cref{fig:power-spectra-resolutions-10mpc} we
first compare results for \ac{FDM} simulations in a \SI{10}{\per\hHubble\Mpc}
box using different resolutions, with grid sizes from $2048^3$ to
$8640^3$. Focusing on the three redshifts of $z = 31$, $z = 10$ and $z = 3$, we see
that the results still appear well converged at the high redshifts of
$z = 31$ and $z = 10$ (although the lowest-resolution simulation with $N^3 = 2048^3$ slightly starts to lag behind at $z = 10$ on smaller scales).
However, this ceases to be true at the lower redshift of
$z = 3$, where the growth of the lower-resolution simulations now clearly trails
behind, and this effect occurs on an extended range of scales,
including fairly large ones, quite unlike in \ac{CDM}, where the impact of
spatial resolution limits is typically constrained to fairly small
scales.
In particular, we see that once the
onset of non-linear evolution is not properly resolved any more in a
simulation, the \ac{FDM} spectra even stop to follow linear growth on
the largest scales, making the simulations appear to be frozen in in
their current state.

Further insights into these numerical effects can also be obtained by
examining how they vary with box size and axion particle mass at a
fixed epoch of $z = 3$. This is shown in
\cref{fig:power-spectra-masses,fig:power-spectra-boxsizes}. Here, the
impact of \cref{eq:velocity-criterion-resolution} is clearly visible:
The resolution requirements become more stringent with larger axion
masses and at fixed grid size, a larger box size directly implies
worse resolution. Thus, at fixed grid size, if axion mass and box
size are scaled by the same factor, the resulting power spectra are
identical except at small scales, which are impacted by the change in
Jeans length with different axion masses. Accordingly, due to
resolution effects and contrary to the physical expectation, for fixed
box size and grid resolution, it is possible to see \emph{more}
small-scale power when the particle mass is lowered, since lower
masses have less stringent resolution requirements.

We finally come to a direct comparison of \ac{FDM} with the
(non-linear) \ac{CDM} power spectrum in
\cref{fig:power-spectra-fdm-vs-cdm}. While both accurately agree with
each other and with linear perturbation theory at large scales and
early times, some differences start to show up as early as
$z ≈ 15$. The non-linear small-scale power enhancement sets in
somewhat more vigorously for \ac{CDM} than for \ac{FDM}. For example,
while mild non-linear amplification is similar at
$k ≈ \SI{10}{\hHubble\per\Mpc}$ in \ac{CDM} and \ac{FDM} for $z = 15$,
this effects extends to smaller scales $k ≈ \SI{100}{\hHubble\per\Mpc}$
in \ac{CDM}, whereas \ac{FDM} still pretty much tracks linear growth
there. From the viewpoint of the overall time evolution, it can thus
be said that the onset of non-linear structure formation is delayed
for \ac{FDM}, and does not proceed strictly in the same bottom-up
fashion as in \ac{CDM}.

Interestingly, \ac{FDM} however eventually catches up in its
non-linear growth and then shows a quite similar overall shape of the
non-linear power spectrum. While \ac{FDM} power is still somewhat
suppressed at $z ≤ 7$ for non-linear scales, the difference to
\ac{CDM} is much smaller than at earlier redshifts. This is true
until a characteristic scale of around
$k ≈ \SI{1000}{\hHubble\per\Mpc}$, where the \ac{FDM} non-linear power
drops significantly below the \ac{CDM} power. This
scale appears to be related to the \q{quantum pressure} effects in \ac{FDM},
and is thus ultimately related to the particle mass.

Curiously, just before the non-linear \ac{FDM} power decays away from
\ac{CDM} towards small scales, it manages to slightly exceed it. This effect could be a reflection of the interference
patterns in the dark matter density field resulting from the wave-like
nature of \ac{FDM}, which generates transient structures of size
$λ_{\mtext{dB}}$ that are clearly visible in high-resolution
images that zoom in on large haloes.\footnote{%
	A similar effect (although in a different cosmological context) has been found in \citet{2020MNRAS.494.2027M}.%
}
These types of order-unity density fluctuations are absent in \ac{CDM}.
If the associated enhancement of
power localized around a characteristic scale $k$ could be measured
by some tracer sufficiently accurately, it could be a tell-tale sign
of \ac{FDM}.

\begin{figure}
	\centering
	\import{res/}{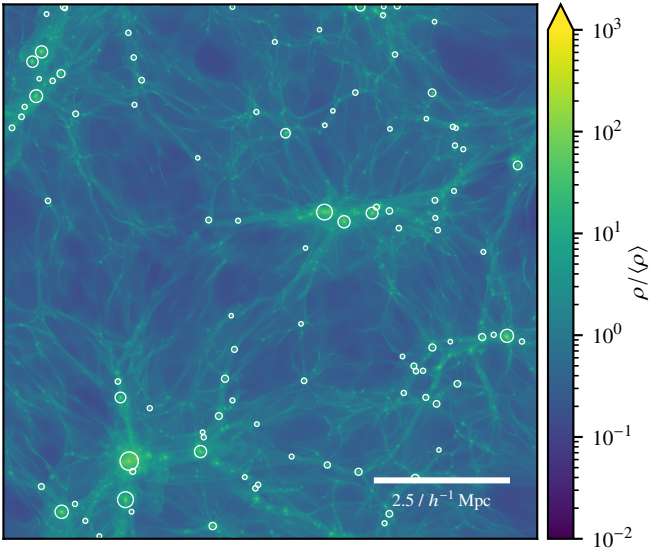}

	\caption{%
		Projected dark matter density at $z = 3$ of a \SI{3}{\per\hHubble\Mpc} (comoving) slab in a high-resolution cosmological box simulation of \acs*{FDM} with box size $L = \SI{10}{\per\hHubble\Mpc}$, \acs*{FDM} mass $m c^2 = \SI{7e-23}{\eV}$, mesh size $N^3 = 8640^3$, and \acs*{CDM} \acsp*{IC}.
		The largest haloes identified with the halo finder are marked with circles whose radii indicate the haloes’ virial radii $R_{200}$.
	}
	\label{fig:halo-finder-slab}
\end{figure}

\begin{figure}
	\newcommand{\plothookfit}{Fit to \cref{eq:hmf-fit} with}%
	\centering
	\import{res/}{mf_dndlnm_m7e-23_mass_function_0.2500.pgf}

	\smallskip

	\import{res/}{mf_dndlnm_m3.5e-23_mass_function_0.2501.pgf}

	\caption{%
		\acs*{HMF} of cosmological \acs*{FDM} and \acs*{CDM} simulations at $z = 3$ with a box size of $L = \SI{10}{\per\hHubble\Mpc}$, \acs*{CDM} \acsp*{IC}, and \acs*{FDM} masses of $mc^2 = \SIlist{7e-23; 3.5e-23}{\eV}$ (top and bottom panels, respectively).
		The \acs*{HMF} derived for \acs*{CDM} by \citet{Sheth1999} is shown for comparison.
		The dotted lines show the fitting function determined by \citet{Schive2016} for the given \acs*{FDM} mass, while the dashed lines show a fit of the data to a similar function (\cref{eq:hmf-fit}) with two free parameters.
		The lower sub-panels show the ratios of the mass functions to the result of the \acs*{CDM} simulation.%
	}
	\label{fig:halo-mass-function}
\end{figure}

\section{Halo mass function}
\label{sec:mass-function}

One of the most important and fundamental outcomes of structure
formation are gravitationally collapsed structures of dark matter,
so-called haloes, which are in turn the sites of baryonic galaxy
formation processes. The abundance of haloes as a function of mass and
epoch is thus fundamental to any cosmological model.

For the \ac{CDM} scenario, the (extended) Press–Schechter formalism
\citep{Press1974,Sheth1999} has proven to be a simple and quite
reliable approach to estimate the \ac{HMF} and its evolution based on
the linear theory power spectrum and the linear growth factor
alone. Comparisons to full $N$-body simulations have both verified the
basic approach and led to the calibration of more accurate empirical
fitting formulae describing the \ac{HMF} \citep{Jenkins2001,Tinker2008,Despali2016}.
As a result, the \ac{HMF} in pure \ac{CDM} models
is now accurately known and understood.

In contrast, the situation for \ac{FDM} is much more murky. While a
few analytic estimates have been published \citep{Marsh2014}, based
loosely on the idea of a Jeans-filtered power spectrum to account for
the \q{quantum pressure}, it is still unclear whether these approaches are
quantitatively reliable. Recently, \citet{Kulkarni2020} have advocated
that a sharp $k$-space filtering in the Press–Schechter formalism with
a variable cut-off may be more adequate for the \ac{FDM} regime. But
their predictions could only be compared to \ac{FDM} mass function
estimates by \citet{Schive2016}, which in turn were based on an
approximate technique of removing \q{spurious} low-mass haloes in a
collisionless $N$-body simulation with truncated initial fluctuation
spectrum, similar to how warm dark matter models are often treated
\citep{Wang2007}. Ultimately, full non-linear simulations of the
Schrödinger–Poisson system are required to obtain quantitatively
reliable results. However, due to the numerical challenges involved in
large-volume simulations of \ac{FDM}, such determinations have not
been obtained thus far.

Our new simulations for the first time allow corresponding
measurements, although the results are not representative of a fully consistent \ac{FDM} cosmology due to the use of \ac{CDM} \acp{IC}.
However, our simulations allow us to demonstrate how strongly the \ac{FDM} dynamics alone impact the \ac{HMF}, and accordingly, to what extent their omission might affect the result.
To this end, we study the \ac{HMF} using the grid-based
halo finder introduced in \cref{sec:halo-finder}, whose operation on our largest simulation is demonstrated in \cref{fig:halo-finder-slab}. In
\cref{fig:halo-mass-function}, we show the obtained mass functions for
the \ac{FDM} and \ac{CDM} cases for a set of example simulations.\footnote{%
	The halo masses are determined using the spherical overdensity definition, i.\,e.\ identifying the densest cell (\ac{FDM}) or the particle with the minimum gravitational	potential (\ac{CDM}) as the centre of a halo, the halo's virial mass $M_{200}$ is defined as the enclosed mass of a sphere around the halo centre with a radius such that the enclosed region has a mean density of $ρ_{200} = 200 ⟨ρ⟩ = 200 Ω_{\mtext{m}} ρ_{\mtext{crit}}$.%
}
As expected from the power spectra, the \ac{HMF} of \ac{FDM} exhibits a
lack of low-mass haloes, but agrees for massive haloes. To obtain
essentially perfect agreement with \ac{CDM} for the most massive haloes,
however, requires sufficiently high resolution in the \ac{FDM}
calculation for the target redshift. As we have seen earlier, the
numerical resolution requirements become ever more stringent towards
lower redshift, and once they start to be compromised, the growth on
large (even still linear) scales becomes damped, which then manifests
itself in halo masses which are biased low. This is clearly seen in
\cref{fig:halo-mass-function} in the comparison of the low- and
high-resolution \ac{FDM} results for the \ac{HMF}.

More interesting, however, is the abundance reduction of low-mass
haloes in \ac{FDM}. Indeed, the delayed onset of structure formation
becomes apparent even more clearly in the \ac{HMF} than in the power
spectrum, with almost no objects present before $z = 7$. At late
times, there is a dearth of low-mass haloes in our results, but the
deficit appears not as strong as the cut-off predicted by
\citet{Kulkarni2020}. Rather, it appears that power transfer to
smaller scales is sufficiently strong in \ac{FDM} that the break in
the \ac{HMF} is weaker than predicted by simplified models based
on \q{quantum pressure}-filtered versions of the Press–Schechter
formalism. This reinforces the notion that simply applying a Jeans
filtering at the linear level and assuming that all smaller scales
never grow and thus can be ignored does not necessarily yield
sufficient quantitative accuracy to predict the non-linear outcome of
\ac{FDM}.
On the other hand, the simulations shown in \cref{fig:halo-mass-function} do not yet include the effect of \acp{IC} appropriate for the \ac{FDM} model, which feature a strong cut-off even in the initial power spectrum (see \cref{sec:fdm-ics}) and are thus expected to result in additional suppression of low-mass haloes.

Also shown in \cref{fig:halo-mass-function} are two fits, given by dotted and dashed lines, in a fashion similar to \citet{Schive2016}.
While the dotted line uses exactly the same function as given in \citet{Schive2016} for the corresponding particle mass $m$, the dashed line is a modified fit of the form
\begin{equation}
	\label{eq:hmf-fit}
	\ml.\frac{\dd n}{\dd M}\mr|_{\mtext{FDM}}\!\!\!(M, z) =
	\ml(1 + \ml(\frac{M}{M_0} \ml(\frac{mc^2}{\SI{e-22}{\eV}}\mr)^{\sfrac{4}{3}} \mr)^α\mr)^{-2.2} \ml.\frac{\dd n}{\dd M}\mr|_{\mtext{CDM}}\!\!\!(M, z),
\end{equation}
where the \citet{Sheth1999} mass function has been used for the \ac{CDM} mass function $(\dd n/\dd M)\big|_{\mtext{CDM}}(M, z)$.
In other words, the mass parameter $M_0$ and the inner exponent $α$ are allowed to vary compared to the values found in \citet{Schive2016} (where they are set to $M_0 = \SI{1.6e10}{\Msunit}$ and $α = -1.1$).
The result shows that while an appropriate fit to our mass function can be obtained using a very similar functional form, the shape changes slightly, and the peak mass $M_0$ shifts to much lower halo masses.
This indicates that either the method used by \citet{Schive2016} underestimates the number of low-mass haloes, or that the inclusion of \ac{FDM}-appropriate \aclp{IC} (as discussed in \cref{sec:fdm-ics}) causes a strong impact on the formation of low-mass haloes.
This question can only be resolved by performing another set of full
\ac{FDM} simulations with self-consistent initial conditions.

\begin{figure}
	\newcommand{\plothooksoliton}{ (\cref{eq:soliton-fit})}%
	\newcommand{\plothookNFW}{ (\cref{eq:nfw-fit})}%
	\centering
	\import{res/}{hp_z3_fdm_halo_profiles.pgf}

	\smallskip

	\import{res/}{hp_z3_cdm_halo_profiles.pgf}

	\caption{%
		\q{Stacked} halo profiles at $z = 3$ for several bins of virial mass $M_{200}$ (given as intervals $[\cdot, \cdot]$) in $L = \SI{10}{\per\hHubble\Mpc}$ cosmological box simulations of \acs*{FDM} ($m c^2 = \SI{7e-23}{\eV}$, $N^3 = 8640^3$) and \acs*{CDM} ($512^3$ particles), using \acs*{CDM} \acsp*{IC}.
		The number of haloes contained in each mass bin is stated in parentheses.
		The top panel shows the results for \acs*{FDM}, while the bottom panel shows those for \acs*{CDM}.
		Thin dashed lines show \acs*{NFW} fits (\cref{eq:nfw-fit}) to the region within the virial radius $R_{200}$, which is indicated with thick lines.
		For \acs*{FDM}, the inner region of the halo has been excluded from the \acs*{NFW} fit and instead been fit to the soliton density profile (\cref{eq:soliton-fit}), as indicated by dot-dashed lines.
		The \acs*{FDM} grid resolution is shown as a vertical dotted line.%
	}
	\label{fig:halo-profiles}
\end{figure}

\section{Halo profiles}
\label{sec:profiles}

The internal density structure of cosmological dark matter haloes is a
further key outcome of non-linear structure formation whose importance
can hardly be overstated. This is because of its crucial influence on
the size, kinematics and morphology of galaxies forming in the dark
matter haloes, as well as on the gravitational lensing strength of
these objects. These properties are of course decisive for the
viability of a cosmological model in the first place. For \acl{CDM} models, the spherically averaged density profiles have
$ρ(r) ∝ r^{-1}$ density cusps at their centres
\citep{Navarro1996}, giving rise to a particular rotation curve shape
of galaxies.

One of the often cited motivations for considering \ac{FDM} is the
so-called cusp–core problem in \ac{LCDM}, combined with the
expectation that \ac{FDM} naturally produces cored density profiles that
may be potentially easier to reconcile with rotation curve data of
certain galaxies.
By considering a spherically-symmetric ground state of \ac{FDM}, one arrives at solutions called solitons, whose density profiles are flat towards their centres.
The production of such solitonic cores has been
predicted on theoretical grounds and has been verified with
simulations of small cosmological volumes \citep{Schive:2014dra}. Here,
we are interested in testing whether we also see them in our
comparatively large-volume simulations, which have however quite
limited ability to resolve the internal structure of individual haloes.

In \cref{fig:halo-profiles} we therefore consider spherically
averaged density profiles, stacked for several different mass ranges,\footnote{%
	The \q{stacking} is performed by determining the mean density across all haloes in a given mass bin for each radial bin.%
}
in our highest-resolution simulation at $z = 3$. The stacking greatly
reduces halo-to-halo scatter and allows us to clearly identify the
mean density profile. For comparison, we also show in the bottom panel
equivalent density profiles for haloes in a corresponding \ac{CDM} $N$-body
simulation. As before, the centres of haloes have been identified as the densest
cell in the \ac{FDM} case, or the particle with the minimum gravitational
potential in \ac{CDM}, while the total halo mass has been determined with
the spherical overdensity definition in both cases. The measurements
of the density profiles include all mass around the halo centres,
i.\,e.\ beyond $R_{200}$, the mean profiles eventually slow their decline
and asymptote to the cosmic mean density.

In the \ac{FDM} case, we clearly see the formation of solitonic cores in
low-mass haloes, as illustrated by the dot-dashed lines, which are
fits to the analytic approximation of the soliton density profile by \citet{Schive2014prl}:\footnote{%
	As before, $ρ_{\mtext{c}}$ refers to the comoving density; $r_{\mtext{c}}$ is the comoving distance from the soliton centre.%
}
\begin{equation}
	\label{eq:soliton-fit}
	ρ_{\mtext{c,soliton}}(r_{\mtext{c}}) =
	\frac{\num{1.9e9}\, a^{-1} \ml(\frac{mc^2}{\SI{e-23}{\eV}}\mr)^{-2} \ml(\frac{r_{\mtext{s}}}{\si{\kpc}}\mr)^{-4}}{\ml(1 + \num{0.091} \ml(\frac{r_{\mtext{c}}}{r_{\mtext{s}}}\mr)^2\mr)^8} \si{\Ms\per\kpc\cubed}
	.
\end{equation}
Note that this profile
depends on the core radius $r_{\mtext{s}}$ only, without further free parameters. Our
simulations thus accurately reproduce earlier work on the shape of the
innermost mean profile, where the \q{quantum pressure} dominates.\footnote{%
	Since we use the mean of many halo density profiles in mass bins of a given width, the physical interpretation of fitting the soliton profile \cref{eq:soliton-fit}, which describes a \emph{single} soliton (with a given mass/radius), to such mean profiles is not necessarily immediately clear.
	The haloes in a given mass bin have a range of (soliton) masses, which are distributed according to the \ac{HMF}.
	However, since the soliton profile is simply flat in the centre and for most of its domain of applicability, such a fit will simply produce a central density corresponding to the \ac{HMF}-weighted mean soliton mass of the binned haloes.
	Although it is not necessarily to be expected that the drop in density beyond the flat central plateau of the stacked profile will be accurately described by the soliton profile due to the varying soliton radii contained in each mass bin, the general agreement with the soliton profile demonstrates the presence of the cores described in previous work in our simulations.%
}
In the outer parts of the haloes, the run of the mean density follows the \ac{NFW}
form \citep{Navarro1996}
\begin{equation}
	\label{eq:nfw-fit}
	ρ_{\mtext{c,NFW}}(r_{\mtext{c}}) =
	\frac{ρ_0}{\frac{r_{\mtext{c}}}{R_{\mtext{s}}} \ml(1 + \frac{r_{\mtext{c}}}{R_{\mtext{s}}}\mr)^2}
\end{equation}
rather well. It is thus tempting to use a combination of a
solitonic core and a \ac{NFW} profile as comprehensive description of the
non-linear density structure of \ac{FDM} haloes, and to use this to forecast
the rotation curve of galaxies and the core size of galaxies as a
function of their virial mass \citep{Burkert2020}. We caution, however,
that this neglects the potentially significant impact of order-unity
density fluctuations of the \ac{FDM} wave function outside the solitonic
core, which are completely lost in the spherical averaging and the
halo stacking. Likewise, it is unclear how severely baryons can impact
the innermost density profile. These baryonic effects are quite
uncertain in \ac{CDM}, but even less is known about them in \ac{FDM}, where they
could potentially be stronger due to the lower dark matter densities
in the centres of haloes. It will thus likely require cosmological
hydrodynamic simulations of galaxy formation in \ac{FDM} to arrive at firm
predictions about rotation curve shapes. Our results here confirm that
such calculations are in principle feasible down to intermediate
redshifts in sufficiently large volumes to allow population studies of
galaxies, especially if the resolution can be made spatially adaptive
\citep[also see e.\,g.][]{2020MNRAS.494.2027M}.

\begin{figure*}
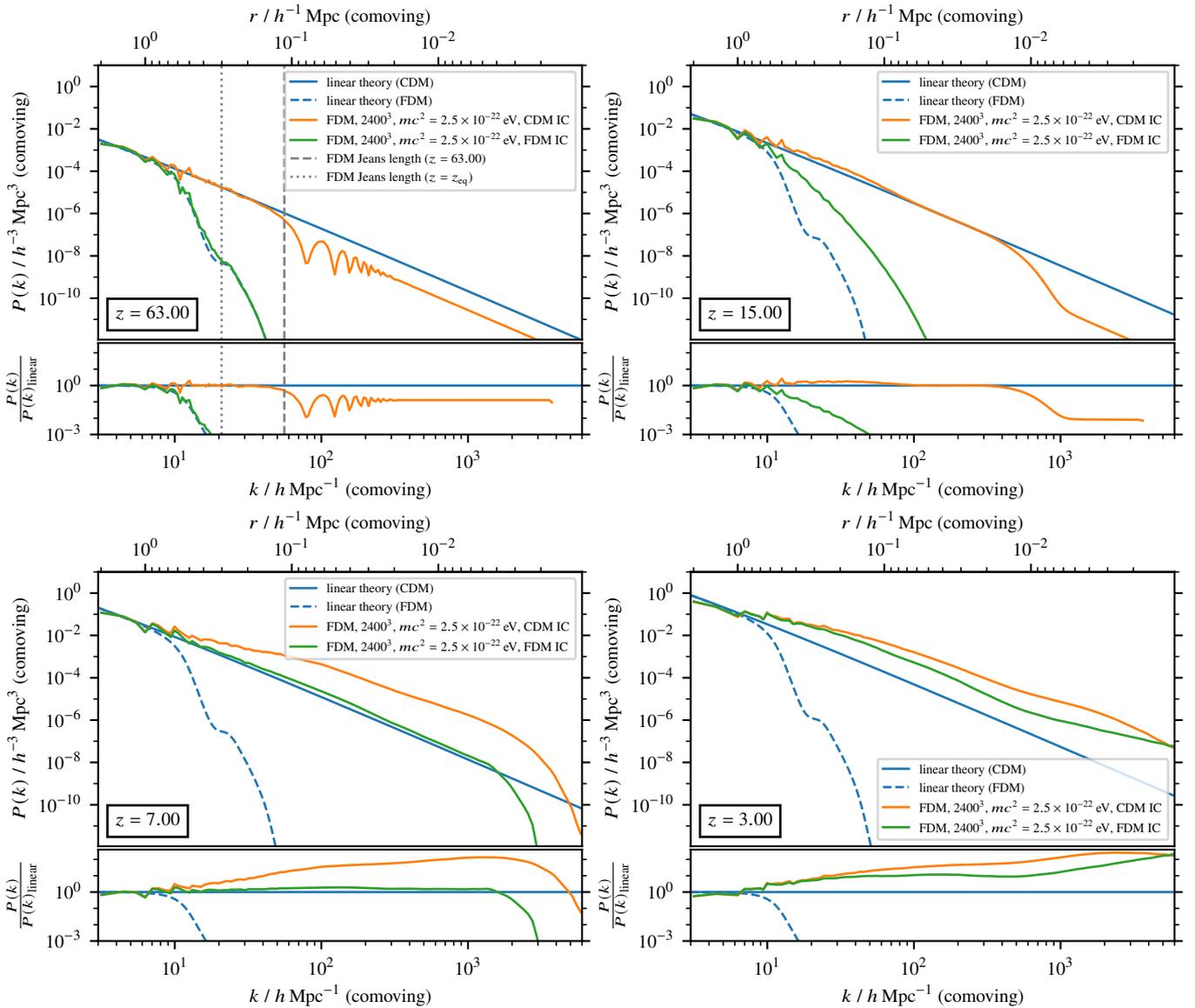

	\centering
	\import{res/psfig_fdmic/}{powerspec_0.0156.pgf}%
	\hfill%
	\hspace*{-0.4in}%
	\import{res/psfig_fdmic/}{powerspec_0.0625.pgf}

	\vspace*{-10ex}

	\import{res/psfig_fdmic/}{powerspec_0.1250.pgf}%
	\hfill%
	\hspace*{-0.4in}%
	\import{res/psfig_fdmic/}{powerspec_0.2500.pgf}

	\caption{%
		Dark matter power spectra at different redshifts for cosmological \acs*{FDM} simulations with both \acs*{FDM} and \acs*{CDM} \acsp*{IC} (\acs*{FDM} mass $m c^2 = \SI{2.5e-22}{\eV}$, box size $L = \SI{2}{\per\hHubble\Mpc}$, mesh size $N^3 = 2400^3$).
		The \acs*{CDM} and \acs*{FDM} initial power spectrum evolved using linear perturbation theory are shown for comparison.
		The lower panels show the ratio of the power spectra to the result from \acs*{CDM} linear theory.
		For $z = 63$, the dashed line additionally indicates the current \acs*{FDM} Jeans scale (\cref{eq:jeans}), while the dotted line indicates the Jeans scale at matter–radiation equality, $z = z_{\mtext{eq}}$.%
	}
	\label{fig:power-spectra-fdm-vs-linear-fdmic}
\end{figure*}

\section{Fuzzy dark matter initial conditions}
\label{sec:fdm-ics}

Finally, we would like to briefly consider the additional source of
differences in \ac{FDM} relative to \ac{CDM} introduced due to modifications of
the initial conditions. Thus far, we had focused for this study on
studying differences between \ac{CDM} and \ac{FDM} resulting just from the
different dynamics induced by the Schrödinger–Poisson system of
equations as opposed to the Vlasov–Poisson system governing
collisionless dynamics of classical particles. To this end, we had
used \emph{identical} initial perturbations, retaining the \ac{CDM} power
spectrum for definiteness. However, a physically consistent model of
a \ac{FDM} cosmology will witness modifications of its initial conditions
relative to \ac{CDM}, particularly in terms of a significant reduction in
small-scale power below a scale set by the \q{quantum pressure}, yielding
something resembling the truncated power spectra of warm dark matter
models. In linear theory, this is determined by the Jeans scale $k_{\mtext{J}}$ (\cref{eq:jeans}),
above which perturbations evolve in an oscillatory rather than growing
fashion.
It is however not clear how scales $k > k_{\mtext{J}}$ react to non-linear
power transfer from larger scales once scales around $k_{\mtext{J}}$
become mildly or fully non-linear. This can only be accurately treated
with explicit (e.\,g.\ spectral) \ac{FDM} simulations like the ones we carry out
here.

In \cref{fig:power-spectra-fdm-vs-linear-fdmic} we show a
comparison of two \ac{FDM} simulations carried out either with \ac{CDM}-like
initial conditions or instead with \acp{IC} predicted by the Boltzmann code
\textcode{axionCAMB} \citep{Hlozek2017} for an axion cosmology with our
adopted particle mass of \SI{2.5e-22}{\eV\per\c\squared} in this
comparison. For definiteness, we chose a \SI{2}{\per\hHubble\Mpc} box size with
$2400^3$ mesh points to represent the wave function; the remaining cosmological parameters are the same as before (\cref{sec:simulations}). We consider the
time evolution down to redshift $z = 3$, and show in each redshift panel
of the plot also the linear theory power spectrum of the corresponding
\ac{CDM} cosmology, for comparison. At $z = 63$, the \ac{FDM} model with \ac{CDM}-like
\acp{IC} follows linear growth accurately only to a scale of order the
Jeans scale set by the \q{quantum pressure}. On much smaller scales, the
growth has completely stalled, while in the transition region
oscillatory behaviour is seen. This is the expected behaviour based on
linear theory, where scales with $k ≫ k_{\mtext{J}}$ should not grow,
while those with $k ≪ k_{\mtext{J}}$ follow the same linear growth as
\ac{CDM}. In comparison, the \ac{FDM} model with self-consistent \acp{IC} shows
negligible small-scale power and only agrees with the \ac{CDM} linear power
on the largest scales that can be followed with this box size.

At redshift $z = 15$, the \ac{FDM} simulation with self-consistent \acp{IC} still
shows a stalled growth at the smallest scales, but scales around
$k_{\mtext{J}}$ have already entered mild non-linear evolution, as
reflected in small excess power relative to the linear theory \ac{CDM}
power spectrum. Apparently, this already couples significantly to
smaller scales, so that scales beyond $k_{\mtext{J}}$ begin to catch up
with the growth, such that in this instance, the power spectrum
happens to agree coincidentally with the linear \ac{CDM} power at around
$k ≈ \SI{200}{\hHubble\per\Mpc}$, while scales
$k ≳ \SI{1000}{\hHubble\per\Mpc}$ are not yet affected and
still prevented from growth. Similarly, the self-consistent \ac{FDM} model
begins to witness power transfer to smaller scales, as evidenced by
the change in shape of its small-scale power spectrum.

By redshift $z = 7$, these trends have greatly accelerated. Now the
non-linear evolution of the \ac{FDM} model started from \ac{CDM}-like initial
conditions has produced a large power excess relative to linear theory
for nearly all $k$ resolved in the simulation, except for the largest
scales (which are still linear), and the smallest scales (which are
still suppressed by \q{quantum pressure}). Interestingly, however, the
model with self-consistent initial conditions is clearly in the
process of catching up to this evolution, with its small-scale power
being progressively filled in by the non-linear evolution on larger
scales. Finally, at redshift $z = 3$, both simulations have become even
closer in their total matter power spectrum. We expect these trends to
continue towards lower redshifts (which are unfortunately inaccessible
with this resolution), so that the discrepancy introduced by starting
from \ac{CDM}-like initial conditions instead of self-consistent \ac{FDM} ones
becomes largely forgotten on small scales, simply because it is
overwhelmed by non-linear evolution. However, there may still be an
important impact on other quantities, such as the \ac{HMF} (which, however, requires a larger box size, as in \cref{sec:mass-function}, to measure).
How large this
is remains to be seen. In principle, excess oscillatory perturbations
on small scales that cannot grow should not give rise to additional
collapsed haloes, but due to the non-linear coupling to larger scales
this may only be approximately true.

\section{Summary and Conclusions}
\label{sec:conclusions}

In this study, we have carried out simulations of fuzzy dark matter
models with the pseudo-spectral method. This can be considered the most
accurate approach to solve the Schrödinger–Poisson system
numerically, without neglecting certain aspects of the temporal
evolution of the wave function. In particular, this can account for
oscillatory, order-unity fluctuations of the local density due to the
quantum-mechanical effects in the axion-like dynamics.

Unfortunately, the numerical resolution requirements to faithfully
follow the \ac{FDM} dynamics are much harder to fulfil than for the
familiar $N$-body techniques applicable in the \ac{CDM} case. Even
large-scale modes require a very fine mesh, otherwise the velocities
appearing in the dynamics cannot properly be represented because then
the spatial variation of the quantum-mechanical phase factor is not
resolved. Yet, making the mesh fine enough to resolve the de Broglie
wavelength $λ_{\mtext{dB}}$ of the largest velocities drives down the time step, which
depends quadratically and not linearly on the spatial resolution. In
practice, this means that one cannot trivially simulate large
cosmological volumes at low resolution in a way similar to the
standard practice in \ac{CDM}. Rather, one is relegated to very small
volumes and to the high-redshift regime, whereas a push to large box
sizes and low redshift quickly becomes extremely expensive, as we
highlighted in this paper.

In this work, we could nevertheless compute the largest-volume simulations with full \ac{FDM} dynamics thus far by using rather large grid sizes of up to
$8640^3$ cells. This allowed us to gain new insights into the
evolution of the non-linear power spectrum in these cosmologies, and
to make the first direct measurements of the \acl{HMF} in
such models, especially with respect to how the \ac{FDM} dynamics affect these observables without starting from an already-suppressed power spectrum in the \acp{IC}.
Our main findings can be summarized as follows:

\begin{itemize}
\item Once sufficient resolution is available,
  the \ac{FDM} power spectrum in our spectral simulations follows
  the \ac{CDM} evolution  very closely on large scales, even
  once highly non-linear evolution has set in on smaller scales.

\item Lacking resolution in pseudo-spectral simulations of \ac{FDM}
  manifests itself in a spurious \q{freezing} of further structure
  formation once the resolution requirements become violated. The
  evolution of the power spectrum then stops, and it remains roughly
  constant with decreasing redshift. This can be detected by
  comparing the power on the largest scales to linear perturbation
  theory or a \ac{CDM} simulation, and observing that the \ac{FDM}
  model falls behind.
  
\item The maximum simulation volume which can be achieved in the
  pseudo-spectral method without violating resolution requirements
  increases for lower particle masses $m$. The computational time
  required for the corresponding maximum box size varies only slowly
  for different values of $m$. Stopping a simulation before
  $z_{\mtext{end}} = 0$ relaxes the resolution requirements, thus allowing
  for larger box sizes, and reduces computational time, but not very
  strongly (the cost approximately scales as
  $∝ z_{\mtext{end}}^{-\sfrac{1}{2}}$ at fixed box size, which
  parallels the scaling of the maximum particle velocity in linear
  perturbation theory).
  
\item On small scales, \ac{FDM} structure formation is suppressed
  compared to \ac{CDM} due to the effects of the \q{quantum pressure},
  which also induces a small-scale cut-off in the initial power
  spectrum of self-consistent cosmological \ac{FDM} models. The onset of
  non-linear structure formation is thus generally delayed in \ac{FDM}
  compared to \ac{CDM}. However, on scales comparable to
  $λ_{\mtext{dB}}$, the difference between \ac{FDM} and \ac{CDM} is
  reduced, and interference patterns on these scales in \ac{FDM} can
  even lead to a small temporary excess of power compared to
  \ac{CDM}. Mildly trans-linear and fully non-linear evolution in
  \ac{FDM} in any case leads to significant power transfer to scales
  smaller than the initial \ac{FDM} Jeans scale, even to the
  extent that oscillatory perturbations on these scales, if present,
  can become completely overwhelmed and buried by the power transfer
  from larger scales.

\item We could, for the first time, measure the \acl{HMF}
  directly from spectral \ac{FDM} simulations. For massive haloes, we
  find the same abundance of haloes as in \ac{CDM}, consistent with
  expectations based on the agreement of their large-scale linear
  power spectra. The formation of the first haloes is however delayed
  for \ac{FDM} with respect to \ac{CDM}, and there are considerably
  fewer low-mass haloes forming compared to \ac{CDM}.
  The low-mass cut-off of the \ac{HMF} we find is less
  pronounced than found by methods which take into account the suppressed \ac{FDM} \acp{IC}, even for those which neglect the non-linear \ac{FDM} dynamics.
  This demonstrates that the \ac{FDM} \acp{IC} have a larger impact on the \ac{HMF} than the dynamics at later times, although the suppression caused by the latter is still sizeable.
  However, since the cut-off appears at much lower halo masses compared to the \ac{FDM} \acp{IC}, the impact of this effect is much more difficult to observe.
  
\item Despite our large grid sizes, our simulations have only limited
  resolving power for the internal structure of dark matter
  haloes. However, we could still clearly detect characteristic
  solitonic cores in spherically averaged density profiles of haloes,
  with a profile shape that matches results of previous
  work well. These cores are particularly large for low-mass haloes, where
  they also substantially reduce the central dark matter densities
  relative to \ac{CDM}.
\end{itemize}

Our results thus highlight that it is, in principle, possible to compute
accurate cosmological simulation results for \ac{FDM} in
representative volumes, albeit at considerable computational cost.
We will apply our implementation to \ac{FDM} initial conditions (similar to \cref{sec:fdm-ics}) in larger volumes in the future, providing physical observables in a fully consistent \ac{FDM} scenario.
However, to conclusively understand the combined dynamics of \ac{FDM} and baryons
in the centres of galaxies, and thus to arrive at reliable predictions
about whether these scenarios can resolve, e.\,g., the cusp–core
tension, will ultimately require to add baryonic physics to these
simulations, and to make them spatially more adaptive inside
individual haloes such that the resolution can be increased there in a
targeted fashion. First attempts in this direction have been carried
out \citep[e.\,g.][]{Mocz2019,Veltmaat2020}, but the associated
numerical challenges are still formidable.

\section*{Acknowledgements}
The authors would like to thank Jowett Chan and Elisa Ferreira for helpful discussions and validation of the numerical method, and Rüdiger Pakmor for guidance concerning the implementation of the simulation code.

\section*{Data availability}
The data underlying this article will be shared on request to the corresponding author.

\renewcommand*{\refname}{REFERENCES}
\bibliographystyle{mnras}
\bibliography{bib}

\begin{thebibliography}{}
\makeatletter
\relax
\def\mn@urlcharsother{\let\do\@makeother \do\$\do\&\do\#\do\^\do\_\do\%\do\~}
\def\mn@doi{\begingroup\mn@urlcharsother \@ifnextchar [ {\mn@doi@}
  {\mn@doi@[]}}
\def\mn@doi@[#1]#2{\def\@tempa{#1}\ifx\@tempa\@empty \href
  {http://dx.doi.org/#2} {doi:#2}\else \href {http://dx.doi.org/#2} {#1}\fi
  \endgroup}
\def\mn@eprint#1#2{\mn@eprint@#1:#2::\@nil}
\def\mn@eprint@arXiv#1{\href {http://arxiv.org/abs/#1} {{\tt arXiv:#1}}}
\def\mn@eprint@dblp#1{\href {http://dblp.uni-trier.de/rec/bibtex/#1.xml}
  {dblp:#1}}
\def\mn@eprint@#1:#2:#3:#4\@nil{\def\@tempa {#1}\def\@tempb {#2}\def\@tempc
  {#3}\ifx \@tempc \@empty \let \@tempc \@tempb \let \@tempb \@tempa \fi \ifx
  \@tempb \@empty \def\@tempb {arXiv}\fi \@ifundefined
  {mn@eprint@\@tempb}{\@tempb:\@tempc}{\expandafter \expandafter \csname
  mn@eprint@\@tempb\endcsname \expandafter{\@tempc}}}

\bibitem[\protect\citeauthoryear{{Boylan-Kolchin}, {Bullock}  \&
  {Kaplinghat}}{{Boylan-Kolchin} et~al.}{2011}]{BoylanKolchin2011}
{Boylan-Kolchin} M.,  {Bullock} J.~S.,   {Kaplinghat} M.,  2011, \mn@doi
  [\mnras] {10.1111/j.1745-3933.2011.01074.x}, \href
  {https://ui.adsabs.harvard.edu/abs/2011MNRAS.415L..40B} {415, L40}

\bibitem[\protect\citeauthoryear{Bull et~al.}{Bull et~al.}{2016}]{Bull:2015stt}
Bull P.,  et~al., 2016, \mn@doi [Phys. Dark Univ.]
  {10.1016/j.dark.2016.02.001}, 12, 56

\bibitem[\protect\citeauthoryear{{Burkert}}{{Burkert}}{2020}]{Burkert2020}
{Burkert} A.,  2020, \mn@doi [\apj] {10.3847/1538-4357/abb242}, \href
  {https://ui.adsabs.harvard.edu/abs/2020ApJ...904..161B} {904, 161}

\bibitem[\protect\citeauthoryear{Del~Popolo \& Le~Delliou}{Del~Popolo \&
  Le~Delliou}{2017}]{DelPopolo:2016emo}
Del~Popolo A.,  Le~Delliou M.,  2017, \mn@doi [Galaxies]
  {10.3390/galaxies5010017}, 5, 17

\bibitem[\protect\citeauthoryear{{Despali}, {Giocoli}, {Angulo}, {Tormen},
  {Sheth}, {Baso}  \& {Moscardini}}{{Despali} et~al.}{2016}]{Despali2016}
{Despali} G.,  {Giocoli} C.,  {Angulo} R.~E.,  {Tormen} G.,  {Sheth} R.~K.,
  {Baso} G.,   {Moscardini} L.,  2016, \mn@doi [\mnras]
  {10.1093/mnras/stv2842}, \href
  {https://ui.adsabs.harvard.edu/abs/2016MNRAS.456.2486D} {456, 2486}

\bibitem[\protect\citeauthoryear{{Edwards}, {Kendall}, {Hotchkiss}  \&
  {Easther}}{{Edwards} et~al.}{2018}]{2018JCAP...10..027E}
{Edwards} F.,  {Kendall} E.,  {Hotchkiss} S.,   {Easther} R.,  2018, \mn@doi
  [\jcap] {10.1088/1475-7516/2018/10/027}, \href
  {https://ui.adsabs.harvard.edu/abs/2018JCAP...10..027E} {2018, 027}

\bibitem[\protect\citeauthoryear{Efstathiou, Sutherland  \& Maddox}{Efstathiou
  et~al.}{1990}]{Efstathiou1990}
Efstathiou G.,  Sutherland W.~J.,   Maddox S.~J.,  1990, \mn@doi [Nature]
  {10.1038/348705a0}, 348, 705

\bibitem[\protect\citeauthoryear{Efstathiou, Bond  \& White}{Efstathiou
  et~al.}{1992}]{Efstathiou1992}
Efstathiou G.,  Bond J.~R.,   White S. D.~M.,  1992, \mn@doi [Monthly Notices
  of the Royal Astronomical Society] {10.1093/mnras/258.1.1P}, 258, 1P

\bibitem[\protect\citeauthoryear{{Ferreira}}{{Ferreira}}{2020}]{Ferreira2020}
{Ferreira} E. G.~M.,  2020, arXiv e-prints, \href
  {https://ui.adsabs.harvard.edu/abs/2020arXiv200503254F} {p. arXiv:2005.03254}

\bibitem[\protect\citeauthoryear{{Frenk} \& {White}}{{Frenk} \&
  {White}}{2012}]{Frenk2012}
{Frenk} C.~S.,  {White} S.~D.~M.,  2012, \mn@doi [Annalen der Physik]
  {10.1002/andp.201200212}, \href
  {https://ui.adsabs.harvard.edu/abs/2012AnP...524..507F} {524, 507}

\bibitem[\protect\citeauthoryear{Frigo \& Johnson}{Frigo \&
  Johnson}{2005}]{Frigo2005}
Frigo M.,  Johnson S.,  2005, \mn@doi [Proceedings of the {IEEE}]
  {10.1109/jproc.2004.840301}, 93, 216

\bibitem[\protect\citeauthoryear{{Garny}, {Konstandin}  \& {Rubira}}{{Garny}
  et~al.}{2020}]{2020JCAP...04..003G}
{Garny} M.,  {Konstandin} T.,   {Rubira} H.,  2020, \mn@doi [\jcap]
  {10.1088/1475-7516/2020/04/003}, \href
  {https://ui.adsabs.harvard.edu/abs/2020JCAP...04..003G} {2020, 003}

\bibitem[\protect\citeauthoryear{{Hlo{\v{z}}ek}, {Marsh}, {Grin}, {Allison},
  {Dunkley}  \& {Calabrese}}{{Hlo{\v{z}}ek} et~al.}{2017}]{Hlozek2017}
{Hlo{\v{z}}ek} R.,  {Marsh} D. J.~E.,  {Grin} D.,  {Allison} R.,  {Dunkley} J.,
    {Calabrese} E.,  2017, \mn@doi [\prd] {10.1103/PhysRevD.95.123511}, \href
  {https://ui.adsabs.harvard.edu/abs/2017PhRvD..95l3511H} {95, 123511}

\bibitem[\protect\citeauthoryear{{Hu}, {Barkana}  \& {Gruzinov}}{{Hu}
  et~al.}{2000}]{Hu2000}
{Hu} W.,  {Barkana} R.,   {Gruzinov} A.,  2000, \mn@doi [\prl]
  {10.1103/PhysRevLett.85.1158}, \href
  {https://ui.adsabs.harvard.edu/abs/2000PhRvL..85.1158H} {85, 1158}

\bibitem[\protect\citeauthoryear{Hui, Ostriker, Tremaine  \& Witten}{Hui
  et~al.}{2017}]{Hui:2016ltb}
Hui L.,  Ostriker J.~P.,  Tremaine S.,   Witten E.,  2017, \mn@doi [Phys. Rev.]
  {10.1103/PhysRevD.95.043541}, D95, 043541

\bibitem[\protect\citeauthoryear{{Jenkins} et~al.,}{{Jenkins}
  et~al.}{1998}]{Jenkins1998}
{Jenkins} A.,  et~al., 1998, \mn@doi [\apj] {10.1086/305615}, \href
  {https://ui.adsabs.harvard.edu/abs/1998ApJ...499...20J} {499, 20}

\bibitem[\protect\citeauthoryear{{Jenkins}, {Frenk}, {White}, {Colberg},
  {Cole}, {Evrard}, {Couchman}  \& {Yoshida}}{{Jenkins}
  et~al.}{2001}]{Jenkins2001}
{Jenkins} A.,  {Frenk} C.~S.,  {White} S.~D.~M.,  {Colberg} J.~M.,  {Cole} S.,
  {Evrard} A.~E.,  {Couchman} H.~M.~P.,   {Yoshida} N.,  2001, \mn@doi [\mnras]
  {10.1046/j.1365-8711.2001.04029.x}, \href
  {https://ui.adsabs.harvard.edu/abs/2001MNRAS.321..372J} {321, 372}

\bibitem[\protect\citeauthoryear{{Kulkarni} \& {Ostriker}}{{Kulkarni} \&
  {Ostriker}}{2020}]{Kulkarni2020}
{Kulkarni} M.,  {Ostriker} J.~P.,  2020, {What is the Halo Mass Function in a
  Fuzzy Dark Matter Cosmology?} (\mn@eprint {arXiv} {2011.02116})

\bibitem[\protect\citeauthoryear{{Lagu{\"e}}, {Bond}, {Hlo{\v{z}}ek}, {Marsh}
  \& {S{\"o}ding}}{{Lagu{\"e}} et~al.}{2021}]{2021MNRAS.504.2391L}
{Lagu{\"e}} A.,  {Bond} J.~R.,  {Hlo{\v{z}}ek} R.,  {Marsh} D. J.~E.,
  {S{\"o}ding} L.,  2021, \mn@doi [\mnras] {10.1093/mnras/stab601}, \href
  {https://ui.adsabs.harvard.edu/abs/2021MNRAS.504.2391L} {504, 2391}

\bibitem[\protect\citeauthoryear{{Li}, {Hui}  \& {Bryan}}{{Li}
  et~al.}{2019}]{2019PhRvD..99f3509L}
{Li} X.,  {Hui} L.,   {Bryan} G.~L.,  2019, \mn@doi [\prd]
  {10.1103/PhysRevD.99.063509}, \href
  {https://ui.adsabs.harvard.edu/abs/2019PhRvD..99f3509L} {99, 063509}

\bibitem[\protect\citeauthoryear{Madelung}{Madelung}{1927}]{Madelung1927}
Madelung E.,  1927, \mn@doi [Zeitschrift für Physik] {10.1007/bf01400372}, 40,
  322

\bibitem[\protect\citeauthoryear{Marsh}{Marsh}{2016}]{Marsh:2015xka}
Marsh D. J.~E.,  2016, \mn@doi [Phys. Rept.] {10.1016/j.physrep.2016.06.005},
  643, 1

\bibitem[\protect\citeauthoryear{{Marsh} \& {Silk}}{{Marsh} \&
  {Silk}}{2014}]{Marsh2014}
{Marsh} D. J.~E.,  {Silk} J.,  2014, \mn@doi [\mnras] {10.1093/mnras/stt2079},
  \href {https://ui.adsabs.harvard.edu/abs/2014MNRAS.437.2652M} {437, 2652}

\bibitem[\protect\citeauthoryear{{Mina}, {Mota}  \& {Winther}}{{Mina}
  et~al.}{2020}]{2020arXiv200704119M}
{Mina} M.,  {Mota} D.~F.,   {Winther} H.~A.,  2020, {Solitons in the dark:
  non-linear structure formation with fuzzy dark matter} (\mn@eprint {arXiv}
  {2007.04119})

\bibitem[\protect\citeauthoryear{{Mocz}, {Vogelsberger}, {Robles}, {Zavala},
  {Boylan-Kolchin}, {Fialkov}  \& {Hernquist}}{{Mocz}
  et~al.}{2017}]{2017MNRAS.471.4559M}
{Mocz} P.,  {Vogelsberger} M.,  {Robles} V.~H.,  {Zavala} J.,  {Boylan-Kolchin}
  M.,  {Fialkov} A.,   {Hernquist} L.,  2017, \mn@doi [\mnras]
  {10.1093/mnras/stx1887}, \href
  {https://ui.adsabs.harvard.edu/abs/2017MNRAS.471.4559M} {471, 4559}

\bibitem[\protect\citeauthoryear{{Mocz}, {Lancaster}, {Fialkov}, {Becerra}  \&
  {Chavanis}}{{Mocz} et~al.}{2018}]{2018PhRvD..97h3519M}
{Mocz} P.,  {Lancaster} L.,  {Fialkov} A.,  {Becerra} F.,   {Chavanis} P.-H.,
  2018, \mn@doi [\prd] {10.1103/PhysRevD.97.083519}, \href
  {https://ui.adsabs.harvard.edu/abs/2018PhRvD..97h3519M} {97, 083519}

\bibitem[\protect\citeauthoryear{{Mocz} et~al.,}{{Mocz}
  et~al.}{2019}]{Mocz2019}
{Mocz} P.,  et~al., 2019, \mn@doi [\prl] {10.1103/PhysRevLett.123.141301},
  \href {https://ui.adsabs.harvard.edu/abs/2019PhRvL.123n1301M} {123, 141301}

\bibitem[\protect\citeauthoryear{{Mocz} et~al.,}{{Mocz}
  et~al.}{2020}]{2020MNRAS.494.2027M}
{Mocz} P.,  et~al., 2020, \mn@doi [\mnras] {10.1093/mnras/staa738}, \href
  {https://ui.adsabs.harvard.edu/abs/2020MNRAS.494.2027M} {494, 2027}

\bibitem[\protect\citeauthoryear{{Navarro}, {Frenk}  \& {White}}{{Navarro}
  et~al.}{1996}]{Navarro1996}
{Navarro} J.~F.,  {Frenk} C.~S.,   {White} S. D.~M.,  1996, \mn@doi [\apj]
  {10.1086/177173}, \href
  {https://ui.adsabs.harvard.edu/abs/1996ApJ...462..563N} {462, 563}

\bibitem[\protect\citeauthoryear{{Nori} \& {Baldi}}{{Nori} \&
  {Baldi}}{2018}]{2018MNRAS.478.3935N}
{Nori} M.,  {Baldi} M.,  2018, \mn@doi [\mnras] {10.1093/mnras/sty1224}, \href
  {https://ui.adsabs.harvard.edu/abs/2018MNRAS.478.3935N} {478, 3935}

\bibitem[\protect\citeauthoryear{{Nori}, {Murgia}, {Ir{\v{s}}i{\v{c}}}, {Baldi}
   \& {Viel}}{{Nori} et~al.}{2019}]{2019MNRAS.482.3227N}
{Nori} M.,  {Murgia} R.,  {Ir{\v{s}}i{\v{c}}} V.,  {Baldi} M.,   {Viel} M.,
  2019, \mn@doi [\mnras] {10.1093/mnras/sty2888}, \href
  {https://ui.adsabs.harvard.edu/abs/2019MNRAS.482.3227N} {482, 3227}

\bibitem[\protect\citeauthoryear{{Press} \& {Schechter}}{{Press} \&
  {Schechter}}{1974}]{Press1974}
{Press} W.~H.,  {Schechter} P.,  1974, \mn@doi [\apj] {10.1086/152650}, \href
  {https://ui.adsabs.harvard.edu/abs/1974ApJ...187..425P} {187, 425}

\bibitem[\protect\citeauthoryear{Schive, Chiueh  \& Broadhurst}{Schive
  et~al.}{2014a}]{Schive:2014dra}
Schive H.-Y.,  Chiueh T.,   Broadhurst T.,  2014a, \mn@doi [Nature Phys.]
  {10.1038/nphys2996}, 10, 496

\bibitem[\protect\citeauthoryear{{Schive}, {Liao}, {Woo}, {Wong}, {Chiueh},
  {Broadhurst}  \& {Hwang}}{{Schive} et~al.}{2014b}]{Schive2014prl}
{Schive} H.-Y.,  {Liao} M.-H.,  {Woo} T.-P.,  {Wong} S.-K.,  {Chiueh} T.,
  {Broadhurst} T.,   {Hwang} W. Y.~P.,  2014b, \mn@doi [\prl]
  {10.1103/PhysRevLett.113.261302}, \href
  {https://ui.adsabs.harvard.edu/abs/2014PhRvL.113z1302S} {113, 261302}

\bibitem[\protect\citeauthoryear{{Schive}, {Chiueh}, {Broadhurst}  \&
  {Huang}}{{Schive} et~al.}{2016}]{Schive2016}
{Schive} H.-Y.,  {Chiueh} T.,  {Broadhurst} T.,   {Huang} K.-W.,  2016, \mn@doi
  [\apj] {10.3847/0004-637X/818/1/89}, \href
  {https://ui.adsabs.harvard.edu/abs/2016ApJ...818...89S} {818, 89}

\bibitem[\protect\citeauthoryear{{Sheth} \& {Tormen}}{{Sheth} \&
  {Tormen}}{1999}]{Sheth1999}
{Sheth} R.~K.,  {Tormen} G.,  1999, \mn@doi [\mnras]
  {10.1046/j.1365-8711.1999.02692.x}, \href
  {https://ui.adsabs.harvard.edu/abs/1999MNRAS.308..119S} {308, 119}

\bibitem[\protect\citeauthoryear{{Springel}}{{Springel}}{2010}]{2010MNRAS.401..791S}
{Springel} V.,  2010, \mn@doi [\mnras] {10.1111/j.1365-2966.2009.15715.x},
  \href {https://ui.adsabs.harvard.edu/abs/2010MNRAS.401..791S} {401, 791}

\bibitem[\protect\citeauthoryear{{Springel}}{{Springel}}{2015}]{2015ascl.soft02003S}
{Springel} V.,  2015, {N-GenIC: Cosmological structure initial conditions}
  (\mn@eprint {ascl} {1502.003})

\bibitem[\protect\citeauthoryear{{Springel} et~al.,}{{Springel}
  et~al.}{2018}]{Springel2018}
{Springel} V.,  et~al., 2018, \mn@doi [\mnras] {10.1093/mnras/stx3304}, \href
  {https://ui.adsabs.harvard.edu/abs/2018MNRAS.475..676S} {475, 676}

\bibitem[\protect\citeauthoryear{{Tinker}, {Kravtsov}, {Klypin}, {Abazajian},
  {Warren}, {Yepes}, {Gottl{\"o}ber}  \& {Holz}}{{Tinker}
  et~al.}{2008}]{Tinker2008}
{Tinker} J.,  {Kravtsov} A.~V.,  {Klypin} A.,  {Abazajian} K.,  {Warren} M.,
  {Yepes} G.,  {Gottl{\"o}ber} S.,   {Holz} D.~E.,  2008, \mn@doi [\apj]
  {10.1086/591439}, \href
  {https://ui.adsabs.harvard.edu/abs/2008ApJ...688..709T} {688, 709}

\bibitem[\protect\citeauthoryear{{Veltmaat} \& {Niemeyer}}{{Veltmaat} \&
  {Niemeyer}}{2016}]{2016PhRvD..94l3523V}
{Veltmaat} J.,  {Niemeyer} J.~C.,  2016, \mn@doi [\prd]
  {10.1103/PhysRevD.94.123523}, \href
  {https://ui.adsabs.harvard.edu/abs/2016PhRvD..94l3523V} {94, 123523}

\bibitem[\protect\citeauthoryear{{Veltmaat}, {Niemeyer}  \&
  {Schwabe}}{{Veltmaat} et~al.}{2018}]{2018PhRvD..98d3509V}
{Veltmaat} J.,  {Niemeyer} J.~C.,   {Schwabe} B.,  2018, \mn@doi [\prd]
  {10.1103/PhysRevD.98.043509}, \href
  {https://ui.adsabs.harvard.edu/abs/2018PhRvD..98d3509V} {98, 043509}

\bibitem[\protect\citeauthoryear{{Veltmaat}, {Schwabe}  \&
  {Niemeyer}}{{Veltmaat} et~al.}{2020}]{Veltmaat2020}
{Veltmaat} J.,  {Schwabe} B.,   {Niemeyer} J.~C.,  2020, \mn@doi [\prd]
  {10.1103/PhysRevD.101.083518}, \href
  {https://ui.adsabs.harvard.edu/abs/2020PhRvD.101h3518V} {101, 083518}

\bibitem[\protect\citeauthoryear{{Wang} \& {White}}{{Wang} \&
  {White}}{2007}]{Wang2007}
{Wang} J.,  {White} S. D.~M.,  2007, \mn@doi [\mnras]
  {10.1111/j.1365-2966.2007.12053.x}, \href
  {https://ui.adsabs.harvard.edu/abs/2007MNRAS.380...93W} {380, 93}

\bibitem[\protect\citeauthoryear{{Weinberg}, {Bullock}, {Governato}, {Kuzio de
  Naray}  \& {Peter}}{{Weinberg} et~al.}{2015}]{2015PNAS..11212249W}
{Weinberg} D.~H.,  {Bullock} J.~S.,  {Governato} F.,  {Kuzio de Naray} R.,
  {Peter} A. H.~G.,  2015, \mn@doi [Proceedings of the National Academy of
  Science] {10.1073/pnas.1308716112}, \href
  {https://ui.adsabs.harvard.edu/abs/2015PNAS..11212249W} {112, 12249}

\bibitem[\protect\citeauthoryear{{Weinberger}, {Springel}  \&
  {Pakmor}}{{Weinberger} et~al.}{2020}]{Weinberger2020}
{Weinberger} R.,  {Springel} V.,   {Pakmor} R.,  2020, \mn@doi [\apjs]
  {10.3847/1538-4365/ab908c}, \href
  {https://ui.adsabs.harvard.edu/abs/2020ApJS..248...32W} {248, 32}

\bibitem[\protect\citeauthoryear{{Widrow} \& {Kaiser}}{{Widrow} \&
  {Kaiser}}{1993}]{1993ApJ...416L..71W}
{Widrow} L.~M.,  {Kaiser} N.,  1993, \mn@doi [\apjl] {10.1086/187073}, \href
  {https://ui.adsabs.harvard.edu/abs/1993ApJ...416L..71W} {416, L71}

\bibitem[\protect\citeauthoryear{{Woo} \& {Chiueh}}{{Woo} \&
  {Chiueh}}{2009}]{2009ApJ...697..850W}
{Woo} T.-P.,  {Chiueh} T.,  2009, \mn@doi [\apj] {10.1088/0004-637X/697/1/850},
  \href {https://ui.adsabs.harvard.edu/abs/2009ApJ...697..850W} {697, 850}

\bibitem[\protect\citeauthoryear{{Zhang}, {Kuo}, {Liu}, {Sming Tsai}, {Cheung}
  \& {Chu}}{{Zhang} et~al.}{2018}]{2018ApJ...863...73Z}
{Zhang} J.,  {Kuo} J.-L.,  {Liu} H.,  {Sming Tsai} Y.-L.,  {Cheung} K.,   {Chu}
  M.-C.,  2018, \mn@doi [\apj] {10.3847/1538-4357/aacf3f}, \href
  {https://ui.adsabs.harvard.edu/abs/2018ApJ...863...73Z} {863, 73}

\bibitem[\protect\citeauthoryear{Zhang, Liu  \& Chu}{Zhang
  et~al.}{2019}]{Zhang:2018ghp}
Zhang J.,  Liu H.,   Chu M.-C.,  2019, \mn@doi [Front. Astron. Space Sci.]
  {10.3389/fspas.2018.00048}, 5, 48

\makeatother
\end{thebibliography}

\bsp
\label{lastpage}
\end{document}